\documentclass[reprint,amsmath,amssymb,aps,nofootinbib]{revtex4-1}

\usepackage{graphicx}
\usepackage{dcolumn}
\usepackage{bm}
\usepackage{amsmath,amssymb,graphicx,bbm,mathrsfs,nicefrac}
\usepackage{tikz}
\usepackage{afterpage}
\usepackage{amsmath}
\usepackage{amssymb}
\usepackage{graphicx,textcomp,float,gensymb,wrapfig, enumitem,comment,dsfont,framed,slashed,appendix,wrapfig,xcolor}
\usepackage{ bbold, multirow }
\usepackage{braket} 
\usepackage{ mathrsfs }
\usepackage[small]{caption}
\usepackage{subcaption}
\usepackage{etoolbox,hyperref,makecell}
\usepackage{feynmp}
\usetikzlibrary{arrows.meta}
\usepackage{empheq}
\usepackage{enumitem}
\usepackage[normalem]{ulem}
\setenumerate[1]{label=(\arabic*)}
\newcommand{\bea}{\begin{eqnarray}}  
\newcommand{\eea}{\end{eqnarray}}

\newcommand{\fref}[1]{Fig.~\ref{#1}}
\newcommand{\eref}[1]{Eq.~(\ref{#1})}

\newcommand{\sref}[1]{Section~\ref{#1}}

\newcommand{\cref}[1]{Chapter~\ref{#1}}
\newcommand{\tref}[1]{Table~\ref{#1}}

\usepackage[utf8]{inputenc}

\begin{document}

\title{
How to search for Mirror Stars with Gaia
}

\author{Aaron Howe}
\email{aaron.howe@mail.utoronto.ca}
\affiliation{Department of Physics, University of Toronto, Canada}

\author{Jack Setford}%
\email{jsetford@physics.utoronto.ca}
\affiliation{Department of Physics, University of Toronto, Canada}

\author{David Curtin}
\email{dcurtin@physics.utoronto.ca}
\affiliation{Department of Physics, University of Toronto, Canada}

\author{Christopher D. Matzner}
\email{matzner@astro.utoronto.ca}
\affiliation{Department of Astronomy \& Astrophysics, University of Toronto, Canada}

\date{\today}

\begin{abstract}

We show for the first time how to conduct a direct search for dark matter using Gaia observations.
Its public astrometric data may contain the signals of \emph{mirror stars}, exotic compact objects made of atomic dark matter with a tiny kinetic mixing between the dark and SM photon.
Mirror stars capture small amounts of interstellar material in their cores, leading to characteristic optical/IR and X-ray emissions. 
We develop the detailed pipeline for conducting a mirror star search using data from Gaia and other stellar catalogues, and demonstrate our methodology by conducting a search for toy mirror stars with a simplified calculation of their optical/IR emissions over a wide range of mirror star and hidden sector parameters. 
We also obtain projected exclusion bounds on the abundance and properties of mirror stars if no candidates are found, demonstrating that Gaia is a new and uniquely powerful probe of atomic dark matter.
Our study provides the blueprint for a realistic mirror star search that includes a more complete treatment of the captured interstellar gas in the future.

\end{abstract}

\pacs{Valid PACS appear here}
\maketitle

\section{Introduction and Motivation}
\label{s.intro}

The Gaia telescope \cite{2016A&A...595A...1G} has already  measured the positions and velocities of well over a billion stars,  ushering in a new era of precision astronomy. 
Gaia's unprecedented boon of astrometric data also has tremendous potential to elucidate the nature of \emph{dark matter}.
The velocity distribution of stars in tidal streams or in the stellar disk and halo provides information on the Milky Way halo gravitational potential \cite{Sanderson:2014apa, Ostdiek:2019gnb} and certain features of the dark matter spatial distribution~\cite{Schutz:2017tfp,Necib:2018igl, Necib:2018iwb}, and investigations that were underway prior to Gaia \cite{1960BAN....15...45O,2015ApJ...814...13M} are now vastly accelerated \cite{Schutz:2017tfp}.  
Such analyses indirectly constrain dark matter via its gravitational effects on stars. However, Gaia's large catalogue of astrometric data also allows for \emph{direct} dark matter searches in more exotic and complex dark sectors. In this paper, we 
develop the methodology to search for the optical and infrared emissions of mirror stars~\cite{Curtin:2019lhm, Curtin:2019ngc} made of atomic dark matter~\cite{Goldberg:1986nk,Kaplan:2009de, Kaplan:2011yj,Cline:2012is,Cline:2013pca,Fan:2013yva,Fan:2013tia,Fan:2013bea,Cyr-Racine:2013fsa,Rosenberg:2017qia,Ghalsasi:2017jna,Gresham:2018anj,Essig:2018pzq,Alvarez:2019nwt,Cline:2021itd,Cyr-Racine:2021alc,Blinov:2021mdk}, using public data from the Gaia satellite and other stellar catalogues. We also show how to obtain exclusions on the abundance and detailed properties of mirror stars in the event that no candidates are observed.

Complex dark sectors are highly motivated. From a bottom-up point of view, it is clear that SM baryonic matter is extremely non-minimal, with many forces and particle species that make up $\mathcal{O}(10\%)$ of the matter in our universe. Therefore, in contrast to the hypothesis of a single collisionless dark matter particle, it is plausible that parts of the remaining matter are equally interesting. 
Atomic dark matter~\cite{Goldberg:1986nk,Kaplan:2009de, Kaplan:2011yj,Cline:2012is,Cline:2013pca,Fan:2013yva,Fan:2013tia,Fan:2013bea,Cyr-Racine:2013fsa,Rosenberg:2017qia,Ghalsasi:2017jna,Gresham:2018anj,Essig:2018pzq,Alvarez:2019nwt,Cline:2021itd,Cyr-Racine:2021alc,Blinov:2021mdk} postulates that dark matter contains at least two states with different masses and opposite charge under a \emph{dark electromagnetism} $U(1)_D$ gauge symmetry.
From a top-down point of view, many theories of Beyond-SM (BSM) particle physics postulate the existence of dark sectors related to the SM via discrete symmetries~\cite{Chacko:2005pe,Barbieri:2005ri,Chacko:2005vw,Chacko:2016hvu,Craig:2016lyx,Chacko:2018vss,Chacko:2021vin,GarciaGarcia:2015pnn,Foot:2002iy,Foot:2003jt,Berezhiani:2003xm,Foot:1999hm,Foot:2000vy,Foot:2003eq, Foot:2004pa}. 
Notably, this includes models of ``neutral naturalness'' such as the Mirror Twin Higgs that address the hierarchy problem of the Higgs boson mass \cite{Chacko:2005pe,Barbieri:2005ri,Chacko:2005vw,Chacko:2016hvu,Craig:2016lyx,Chacko:2018vss,Chacko:2021vin,GarciaGarcia:2015pnn} and could resolve outstanding anomalies in precision cosmology, see~\cite{Bansal:2021dfh} (and also~\cite{Cyr-Racine:2021alc,Blinov:2021mdk}).
This can result in a fraction of dark matter being made up of dark baryons, with dark protons, dark electrons, dark photons, and even dark nuclear forces that are similar to their SM counterparts but with possibly different values for masses and interaction strengths.

The observational consequences of this \emph{dark complexity} are striking. %
A $\lesssim 10\%$ subcomponent of dark matter that is made up of various dark particle species with sizeable self-interactions would be compatible with cosmological and self-interaction bounds~\cite{Fan:2013yva,Cyr-Racine:2013fsa,Chacko:2018vss}. 
Just as Standard Model (SM) particles like protons and electrons interact via electromagnetism, which allows baryonic gas to cool and form structure, these dark particles could interact via dark electromagnetism, emitting (to us) invisible dark photons. The resulting dissipation leads to the formation of dark structure, such as a dark disk or dark microhalos~\cite{Fan:2013yva,Fan:2013tia,Ghalsasi:2017jna}, and mirror stars~\cite{Foot:1999hm, Foot:2000vy, Berezhiani:2005vv, Curtin:2019lhm, Curtin:2019ngc}.

It has recently been shown~\cite{Curtin:2019lhm, Curtin:2019ngc} that mirror stars can produce electromagnetic signals that may be visible in optical/IR and X-ray telescopes.
This is possible if the dark photon $\gamma_D$ and the SM photon $\gamma$ have a small \emph{kinetic mixing} in the Lagrangian $\mathcal{L} \supset \epsilon F_{\mu \nu} F^{\mu \nu}_D$~\cite{Holdom:1985ag}.
This small renormalizable operator is readily generated from many possible loop-processes in the full UV-theory, with cosmological constraints~\cite{Vogel:2013raa} and estimates of gravitational effects~\cite{Gherghetta:2019coi} motivating $10^{-13} \lesssim \epsilon \lesssim 10^{-9}$.
Such a tiny interaction has negligible effects on cosmology, structure formation or astrophysics, but it does result in SM matter (dark sector matter) acquiring a ``dark milli-charge'' (SM milli-charge).
As a result, mirror stars in our galaxy would capture a small amount of regular gas from the interstellar medium, which sinks to the star's center and forms a ``SM nugget'' that siphons off small amounts of energy and re-radiates it as SM photons at optical/IR and X-ray frequencies. 
Mirror stars  would therefore look  to a photometric survey like hot white dwarfs  that are unusually dim and  host unusual X-ray emission. 
Gaia's large dataset and ability to determine the absolute magnitude of stars  makes it the ideal instrument to identify promising mirror star candidates  in the local environment of the Sun.

We adopt here the simplified models of Ref.~\cite{Curtin:2019lhm} for the optical/IR emissions of the SM nugget, and use Gaia's public data to identify anomalously dim white-dwarf like objects that could be mirror stars.  We find a few candidates, which we cross-match to other catalogues in order to assemble rudimentary spectral data from a variety of observed pass-bands. This allows the mirror star candidates to be compared to other white dwarfs, and we (unfortunately) conclude that these objects are most likely just dim and possibly dust-reddened  white dwarfs.

Our current mirror star search analysis must be regarded as provisional, since the simplified optical/IR emission models of~\cite{Curtin:2019lhm,Curtin:2019ngc} only include bremsstrahlung and thermal emission of captured hydrogen and helium. Other processes and captured elements will will significantly affect the shape, though probably not the absolute magnitude, of the optical/IR emission spectrum. It also does not include a detailed treatment of the SM nugget's structure in the case where it is optically thick, which will affect its surface temperature. We intend to conduct a realistic mirror star search with a more complete treatment of the SM nugget in a future work. The methodology we develop in this paper to account for the wide range of mirror star and hidden sector parameters can then be applied to this more realistic SM calculation verbatim.

Finally, we demonstrate how the non-observation of mirror stars can be used to set constraints on the abundance and properties of mirror stars in our immediate stellar neighborhood.\footnote{Since the sensitivity depends mostly on the bolometric magnitude of mirror stars in SM photons, the constraints we derive are robust with respect to the approximations made in the simplified emissions model of the captured SM nugget.}
In general, this depends on the mirror star mass distribution, which is unknown and currently impossible to predict. Fortunately, this task can be simplified by recognising that the most important feature of the mirror star distribution for the purpose of setting constraints is just the mass of the lightest visible mirror stars.
This focuses future work on understanding the relationship between the fundamental Lagrangian of the dark sector and the properties of ``mirror red dwarfs'', i.e. the lightest mirror stars that can form and (if the dark sector has nuclear forces) sustain dark nuclear fusion. 

Our work is highly complementary to other proposed mirror star searches using microlensing surveys~\cite{Winch:2020cju} and gravitational waves~\cite{Hippert:2021fch}, since it probes the size of dark photon kinetic mixing and also provides information about the local mirror star density as opposed to their distribution on galactic or even cosmological scales. 

This paper is structured as follows. In \sref{s.MSproperties} we briefly review how mirror stars produce optical/IR signals.
In \sref{s.MSsearch} we develop our methodology to search for mirror stars in Gaia and other stellar catalogue data, and apply it by conducting a toy search using the simplified treatment for their optical/IR emissions.
\sref{s.MSconstraints} explains how we use non-observation of mirror stars to place constraints on their abundance and properties. In  \sref{s.conclusion} we discuss our results and place them in the context of other mirror star searches.

\section{Mirror Star Properties}
\label{s.MSproperties}

We assume that part of dark matter is composed of atomic dark matter, with dark protons of mass $m_{p'}$ (or, more generally, dark nuclei $N'_i$ of mass $m_{N'_i}$), dark electrons of mass $m_{e'}$, and dark QED with interaction strength $\alpha_D$. This is a simple but sufficient stand-in for more complicated dark sectors in models like the Twin Higgs. Dark atoms will form mirror stars in our galaxy if they can cool efficiently, though their abundance and spatial distribution as a function of the atomic dark matter fraction and other parameters is not known. If the dark sector also features an analogue of strong interactions, then mirror stars will initiate dark nuclear fusion in their core to shine in dark photons for extended periods of time, a phenomenon expected to be robust even for physical constants very different from the SM~\cite{Adams08}. 

The crucial mechanism that leads to Mirror Stars having a detectable signature is the fact that they generically capture Standard Model material from the interstellar medium (ISM) \cite{Curtin:2019lhm,Curtin:2019ngc} if  the dark and SM photons have kinetic mixing  $\epsilon F_{\mu \nu} F_D^{\mu \nu}$ in the Lagrangian. The interaction cross section between regular and mirror matter is proportional to $\epsilon^2$, since mirror matter acquires a millicharge under the SM photon. 
Thus, particles charged under mirror electromagnetism inherit a small interaction with particles charged under regular electromagnetism. This allows for scattering of incoming hydrogen and helium atoms/nuclei off the mirror stellar material, which can lead to capture and, over the course of the mirror star's lifetime, significant accumulation of regular matter in the mirror star's core.\footnote{Note that the reverse process is also expected to occur: see \cite{Curtin:2020tkm}, where the accumulation of dark baryons in white dwarfs is used to place constraints on the local density of atomic dark matter.}
In this section, we qualitatively review how SM matter is captured in mirror stars to form a SM nugget, how this nugget is heated up via interactions with the mirror star core, and how it radiates thermal and X-ray SM photons. For details, see the full discussion in~\cite{Curtin:2019lhm,Curtin:2019ngc}.

There are two distinct capture mechanisms for the incoming SM material: \emph{mirror-capture}, which involves the $\epsilon^2$-suppressed scattering off mirror stellar nuclei; and \emph{self-capture}, which involves the much more efficient scattering off already-captured SM matter. The total mirror-capture rate will be proportional to the total number of targets (i.e.\ dark nuclei) in the mirror star, while self capture is generally so efficient that it is bounded by the geometric limit given by the physical size of the nugget.  Gravitational focusing is taken into account for both mechanisms.\footnote{In~\cite{Curtin:2019lhm,Curtin:2019ngc}, the thermal motion of mirror nuclei was neglected in the computation of self-capture. Taking this thermal motion into account may dramatically increase the mirror-capture rate~\cite{Gaidau:2021vyr,DeRocco:2022rze} and lead to an earlier switch-over to geometric self-capture, which would significantly enhance the mirror star signal for scenarios with low mirror-capture rates. Since capture in the most visible mirror star candidates is already dominated by self-capture, this would not significantly change the boundaries of the mirror star signal regions we derive in \sref{s.MSsearch}, but it does make the constraints we derive for very small values of the kinetic mixing $\epsilon$ conservative.}

It is important to be able to estimate the equilibrium temperature of the SM nugget, since this affects the self-capture rate (because the temperature determines the virial size of the nugget which in turn determines its geometric capture limit), and also because the nugget temperature determines the amount and characteristic frequency of the emitted radiation and hence its  signature in Gaia data. 
In \cite{Curtin:2019lhm} we presented simple estimates for the heating and cooling rate of the SM nugget composed of hydrogen and helium, which allow one to solve for the equilibrium nugget temperature.
These estimates are accurate at the order-of-magnitude level but computationally much faster than the more careful calculations presented in~\cite{Curtin:2019ngc}.
We use these simplified calculation to demonstrate our mirror star search methodology, since this makes it easy to consider large samples of mirror stars with widely varying properties and understand their signatures in generality.

The SM nugget is very small compared to the mirror star. For Sun-like mirror stars, nugget radii of $\mathcal{O}(10^3 \mathrm{km})$ are typical~\cite{Curtin:2019lhm}. We therefore regard the nugget as being entirely contained within the mirror star core. As a result, there are two main energy transfer mechanisms from the mirror star to the SM nugget: 
1) heating via collisions between mirror nuclei and SM nuclei and 2) mirror X-ray conversion. Collisional heating proceeds via the same interaction as mirror-capture -- photon portal induced Rutherford scattering. X-ray conversion refers to the fact that in the presence of both visible and mirror matter, a mirror photon can elastically convert to a Standard Model photon via a Thomson-scattering-like process. These converted photons will have energies characteristic of the Mirror Star's core temperature, thus once converted can be a significant source of heating for X-ray opaque SM nuggets. Some of these photons escape, and their detection in X-ray observations would also provide a direct window into the mirror star's interior, serving both as  a smoking-gun signal of the mirror star's dark matter nature and a detailed probe of dark sector micro-physics.

Cooling of the nugget, on the other hand, is very efficient (involving only Standard Model processes), and tends to result in the nugget being significantly cooler than the core temperature of the Mirror Star. Thus the nugget acts as a constant heat sink from the Mirror Star core, drawing energy from the mirror stellar material which is then efficiently radiated away in SM photons that pass through the mirror star material unobstructed.
(This has negligible effect on mirror star evolution for the range of their properties that we consider.)
In the emission models of  \cite{Curtin:2019ngc, Curtin:2019ngc}, the nugget temperature  self-regulates to $\mathcal O(10^4\,\textrm{K})$ because the free-free
cooling process, being rapidly more efficient with increasing ionization, is very inefficient below the ionization threshold.  This model admits both optically thick and optically thin solutions, with optically thick nuggets assumed to be isothermal for simplicity.

It is important to note that, since the internal structure, surface properties, and emitted  optical/IR spectrum of a nugget are determined by the physics 
of captured \emph{Standard Model} interstellar gas,
the properties of the mirror star signal that we can observe are set in large part by Standard Model physics. The 
observable properties of mirror stars will therefore be
robust over a wide range of mirror stellar physics parameters, allowing for telescope surveys to probe wide regions of BSM parameter space.

In a future work, we intend to address the details of SM nuggets' hydrostatic structure and thermal stability in the context of additional  emission and absorption processes, such as bound-free and bound-bound H$^-$ opacity and the collisionally-induced absorption of H$_2$, among others such as the effect of captured metals (see \cite{Freedman14_Opacities}).  We anticipate that these additional considerations will imply that the transition from optically thin to optically thick nugget occurs at somewhat lower density than predicted in~\cite{Curtin:2019ngc, Curtin:2019ngc}, and that optically thick SM nuggets are larger in radius and cooler in surface emission.
The additional emission processes may also shift the colour of optically thin nuggets.

\section{Search for Toy Mirror Stars}
\label{s.MSsearch}

We now demonstrate how to conduct a search for mirror stars using Gaia and other public stellar catalogue data. The main challenge is taking into account the wide range of possible hidden sector micro-physics parameters, and their currently unknown relationship to the detailed properties of the resulting mirror stars. 
Fortunately, since the optical/IR signals are mostly determined by SM physics, mirror stars tend to show up in well-defined and distinct regions of the Hertzsprung-Russell (HR) colour-magnitude diagram, which can be used to define a mirror star search. We demonstrate this principle by developing a full mirror star search pipeline using the simplified emissions model of~\cite{Curtin:2019ngc}, essentially conducting a particle phenomenology-style BSM search over a large signal parameter space using astrophysics data. This will make it straightforward to conduct a realistic mirror star search with a more complete calculation of the SM nugget properties in the future.

\subsection{Defining a Mirror Star Signal Region}

In principle, it would be possible to determine the range of allowed mirror star properties from the atomic dark matter Lagrangian, but in practice, this is difficult and uncertain in generality. It is more useful to parameterize mirror stars in terms of parameters that directly determine their signature:
\begin{itemize}
    \item total mirror star mass $M_{MS}$,
    \item  radius $r_{MS}$,
    \item core density $\rho_\mathrm{core}$, 
    \item core temperature $T_\mathrm{core}$,
    \item  age $\tau_{MS}$, 
    \item  dark nucleus mass  $m_{N'}$. (If there are multiple dark nuclear species, we take this to be the average.)
\end{itemize}
The mass of the mirror electron, provided it is significantly less than the mirror nuclei, does not affect our estimate for the signature. 
This simple mirror star model is sufficient to estimate the optical/IR emission spectrum because (1) only the total number of targets is needed to calculate mirror-capture, not their spatial distribution, and (2) the virial radius SM nugget will generally be small relative to the size of the star, so we can estimate the heating rate using only the temperature and density at the Mirror Star's core.
This parameterization allows us to factorize the problem of understanding mirror star signatures from the problem of understanding mirror stellar physics.

To illustrate the various possible mirror star signals, we perform a scan by randomly varying each of the above parameters on a log-scale by 2 orders of magnitude both above and below their values for the sun, except for the constituent mass, which we vary by just one order of magnitude above and below the mass of the proton, and the age, which we require to be less than the age of the universe.
We also vary the kinetic mixing parameter and dark QED coupling $\epsilon\sqrt{\alpha_{D}/\alpha_{em}}$ in the range $10^{-10}-10^{-13}$.
The output spectrum of each mirror star's SM nugget can then be convolved with the Gaia passband functions~\cite{2018A&A...616A...4E} to compute its position in the Gaia HR diagram, assuming Gaia could measure the mirror star's parallax with sufficient accuracy to determine its absolute magnitude. 

\begin{figure}
   \includegraphics[scale=0.45]{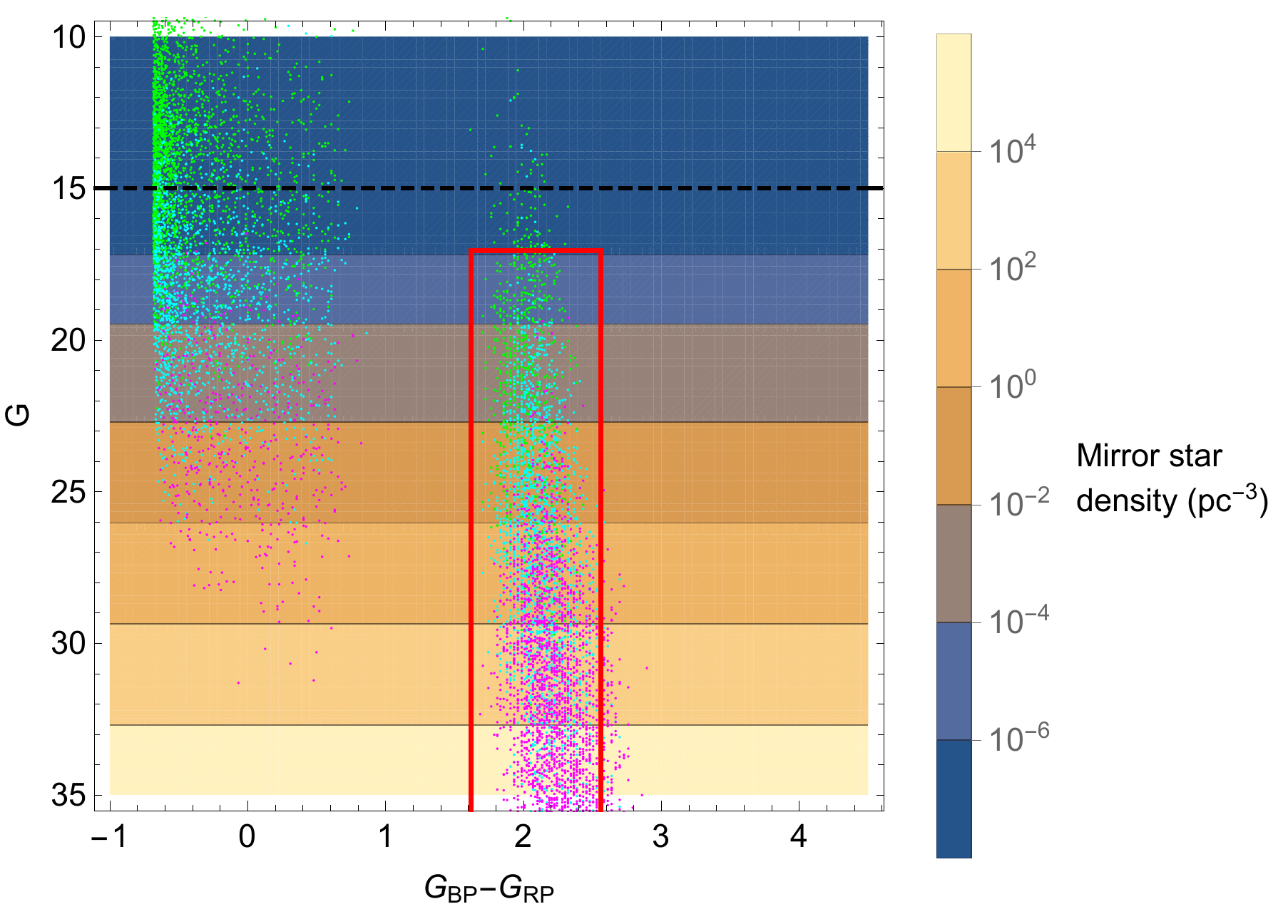}
    \caption{Distribution of our simulated mirror star sample in Gaia's HR diagram. Green dots: $10^{-10} \leq \epsilon \leq 10^{-11}$. Cyan dots: $10^{-11} \leq \epsilon \leq 10^{-12}$. Magenta dots: $10^{-12} \leq \epsilon \leq 10^{-13}$. 
    The red box is the signal region for our mirror star search in \sref{s.MSsearch}. 
    Horizontal contours indicate the mirror star number density that corresponds to 4 expected mirror star observations in Gaia, corresponding to the abundance that could be excluded by a non-observation in Gaia. 
    The dashed black line indicates the magnitude for which the maximum distance for detection is 100 pc. 
    }
    \label{fig:HR_density}
\end{figure}

The results of this scan are shown in Fig.~\ref{fig:HR_density}, which is an HR-diagram in Gaia's passband magnitude coordinate space. The vertical axis is absolute magnitude, i.e. flux measurement in the $G$ (broad) passband, normalized via the parallax measurement. The horizontal axis is the difference between the $G_{BP}$ (bluer) and $G_{RP}$ (redder) passbands, indicating colour or temperature (bluer and hotter to the left). Each mirror star in our sample corresponds to a dot in this plane. 
Horizontal contours indicate the mirror star number density for a given absolute magnitude of its SM nugget that corresponds to 4 expected mirror star detections above Gaia's relative magnitude threshold of 20. 
It therefore corresponds to the mirror star number density that a non-observation at Gaia could exclude at $95\%$-confidence-level. 
Of course, for a given mirror star mass, the mirror star number density is already constrained just by measurements of the local DM mass density~\cite{Bovy:2012tw,Benito:2019ngh}. 
We will discuss new constraints from Gaia in more detail in \sref{s.MSconstraints}.

Fig.~\ref{fig:HR_density} makes it clear that there are  two distinct signal regions. These correspond to the regimes in which the nugget of SM material is optically thin (``cooler'' cluster of points on the right) or optically thick (``hotter'' cluster of points on the left). 
Our criteria for making the optically thin/thick distinction is to compare the size of the nugget to the path length of a photon with energies characteristic of the nugget's average temperature.
Since the properties of the stars in our scan are varying wildly, and the interaction strength between mirror and regular matter is also varying between several orders of magnitude, it is natural to expect that the amount of captured material also varies dramatically. Thus the photon mean free path can range from a small fraction of the nugget's total size to greater than the size of the star itself. Optically thin nuggets can cool more efficiently, since any photon emitted throughout the nugget's volume can escape; optically thick nuggets cool only from their surface. This results in optically thick nuggets having a higher average temperature than optically thin nuggets. 
As discussed above, a full calculation of the optically thick nuggets' properties is still outstanding and would have to take radiative and convective heat transport into account. For the present analysis, we follow~\cite{Curtin:2019lhm, Curtin:2019ngc} assume it to be isothermal for simplicity. 
This may exaggerate the separation between optically thick and thin nuggets in the HR diagram, but is sufficient for the purposes of this toy demonstration study.

Another feature that leads to the separation between these two regions on the HR diagram is the fact that the spectral shape of an optically thin nugget is determined primarily by the spectrum of emitted bremsstrahlung radiation, which is flat at low frequencies. On the other hand, for optically thick nuggets low frequencies are more easily absorbed and the spectrum approaches that of blackbody surface radiation. (This is also responsible for the bunching of points to the left, as the difference between $BP$ and $RP$ passbands approaches a constant for very hot blackbodies.) 
A more realistic nugget calculation will include additional emission processes which may shift the colour of optically thin nuggets, but this will likely preserve the separation from optically thick nuggets in the HR diagram.

Of course the transition from optically thin to optically thick should be gradual during the Mirror Star's evolution, as more matter is accumulated and the nugget density increases. However, by taking into account the full frequency and temperature-dependence of optical depth, as well as the known accumulation rate and hence temperature evolution of the nugget, we have verified that the crossover is very fast compared to the lifetime of the star, so we are justified in not modelling the crossover in detail for our scans. 

In conclusion, our calculation shows that mirror stars are likely to show up in one of two distinct regions of the HR diagram. This informs mirror searches using Gaia data.

\begin{figure*}
\begin{tabular}{ccc}
    \includegraphics[scale=0.3]{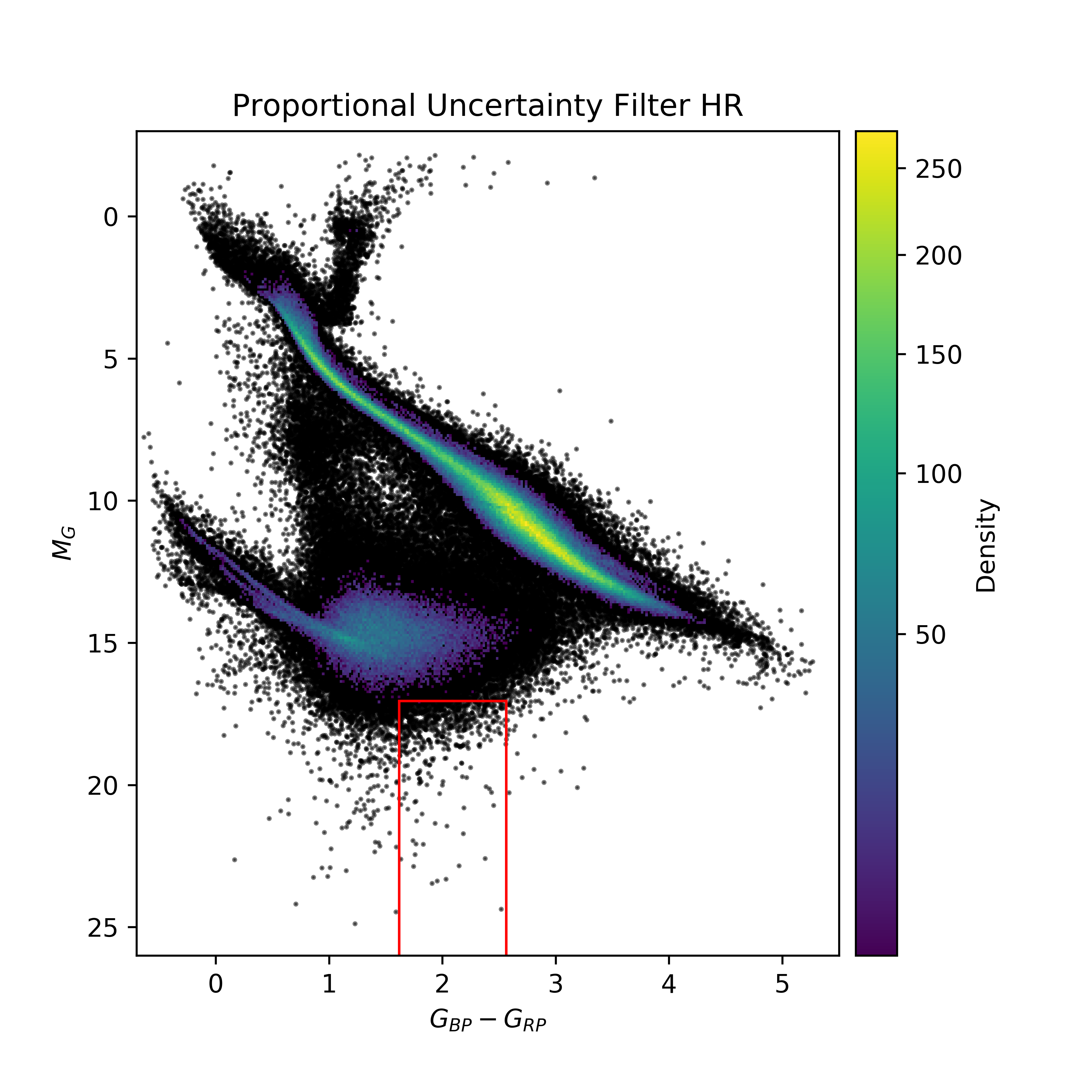} &
    \includegraphics[scale=0.3]{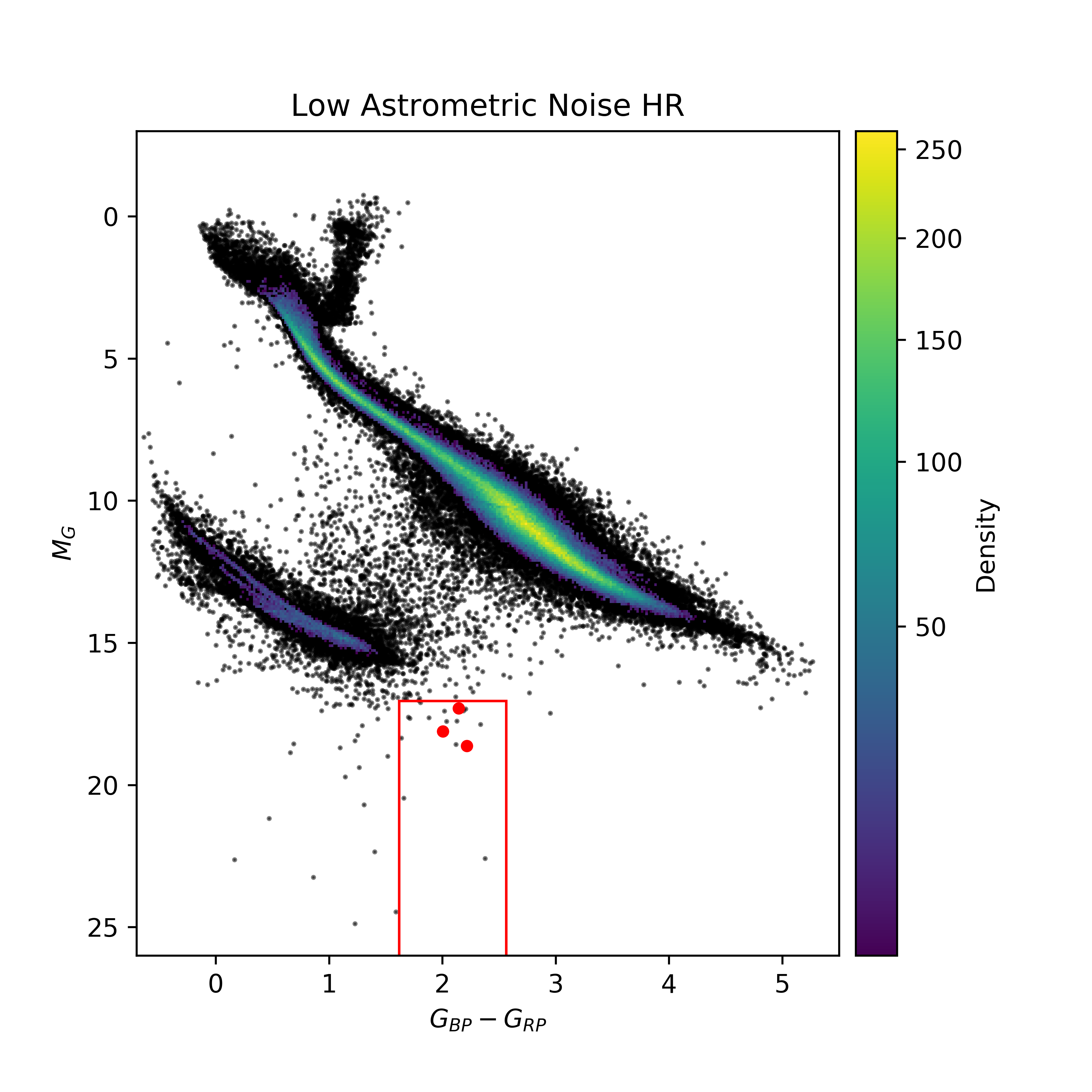} &
    \includegraphics[scale=0.3]{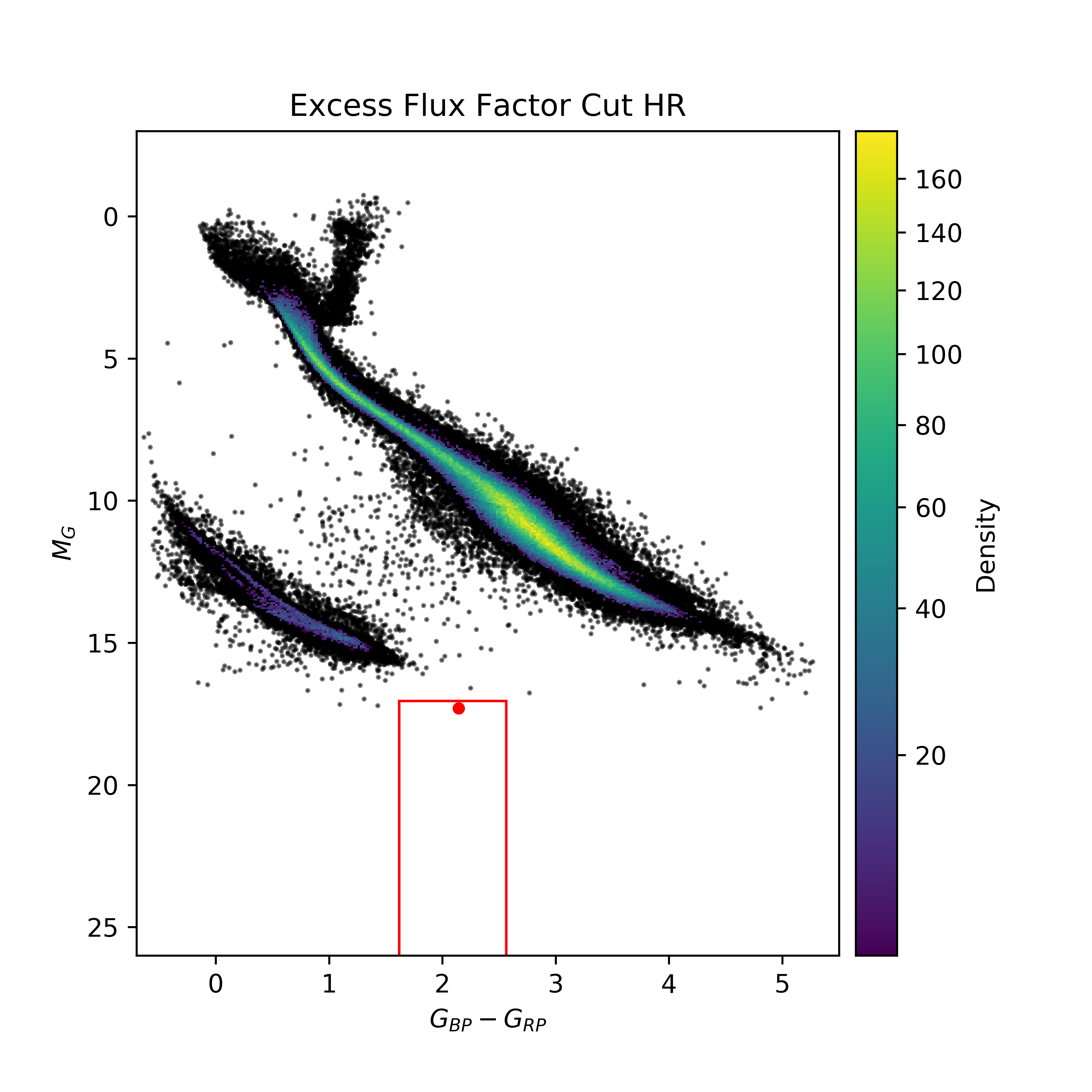}
    \\
    \includegraphics[scale=0.3]{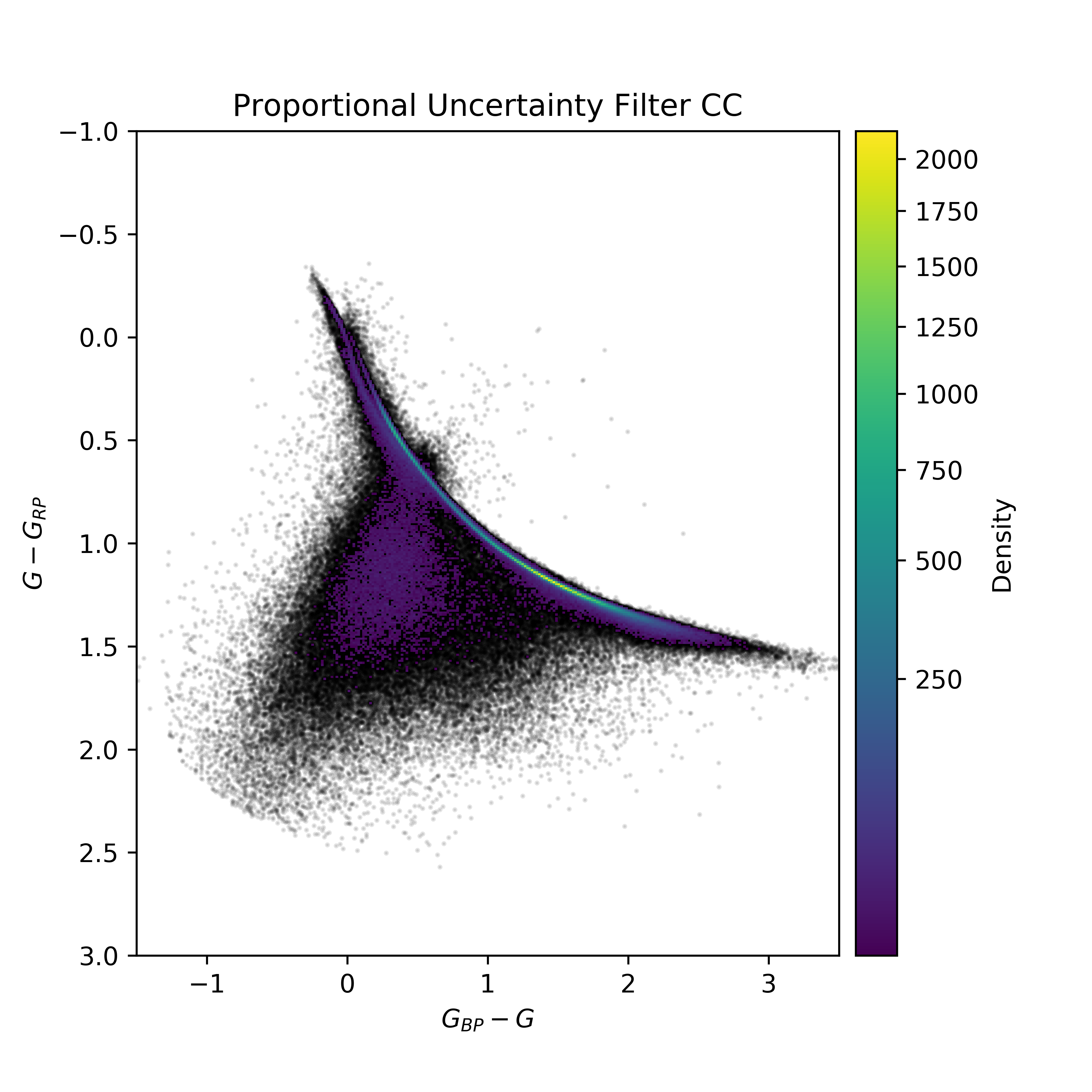}  &
    \includegraphics[scale=0.3]{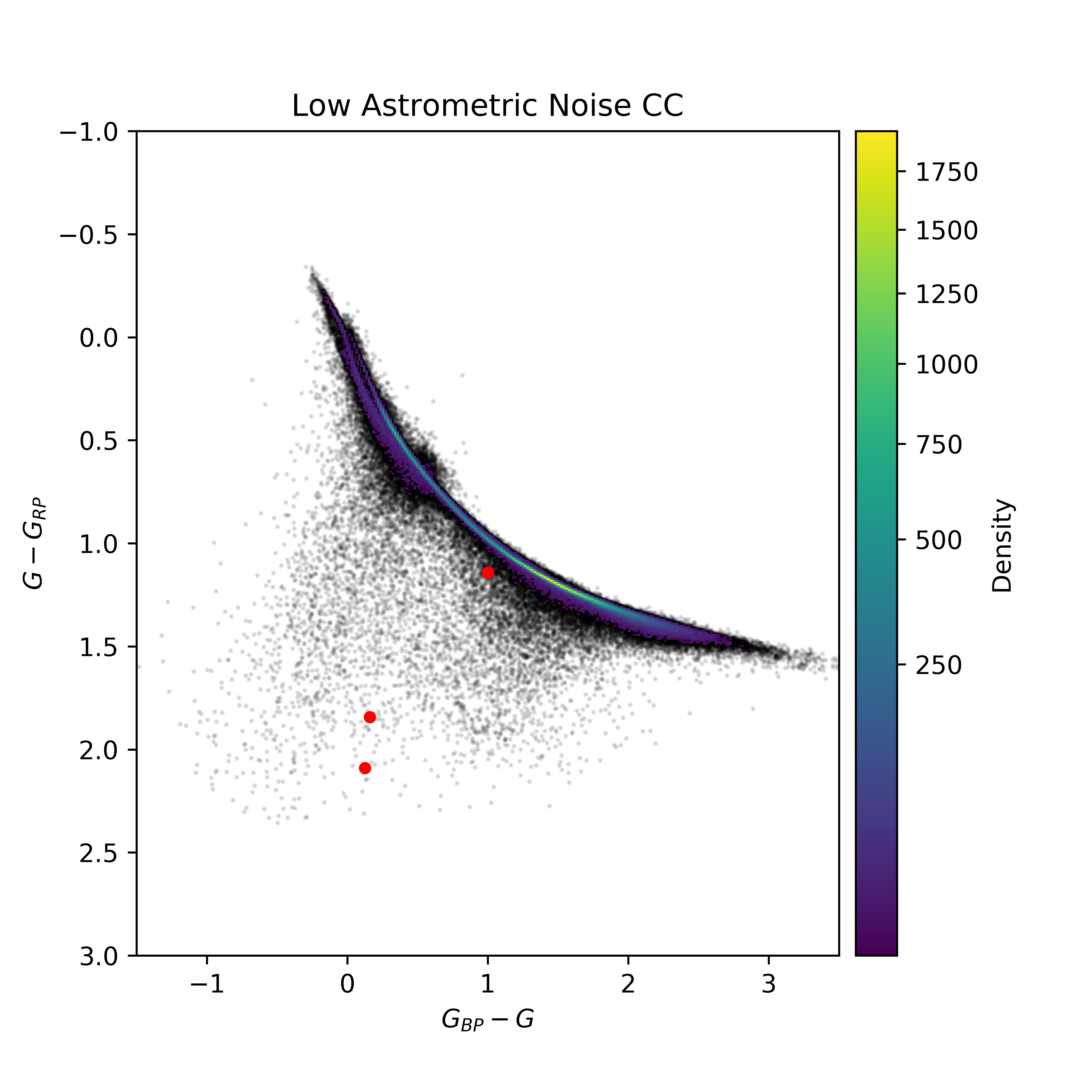} &
    \includegraphics[scale=0.3]{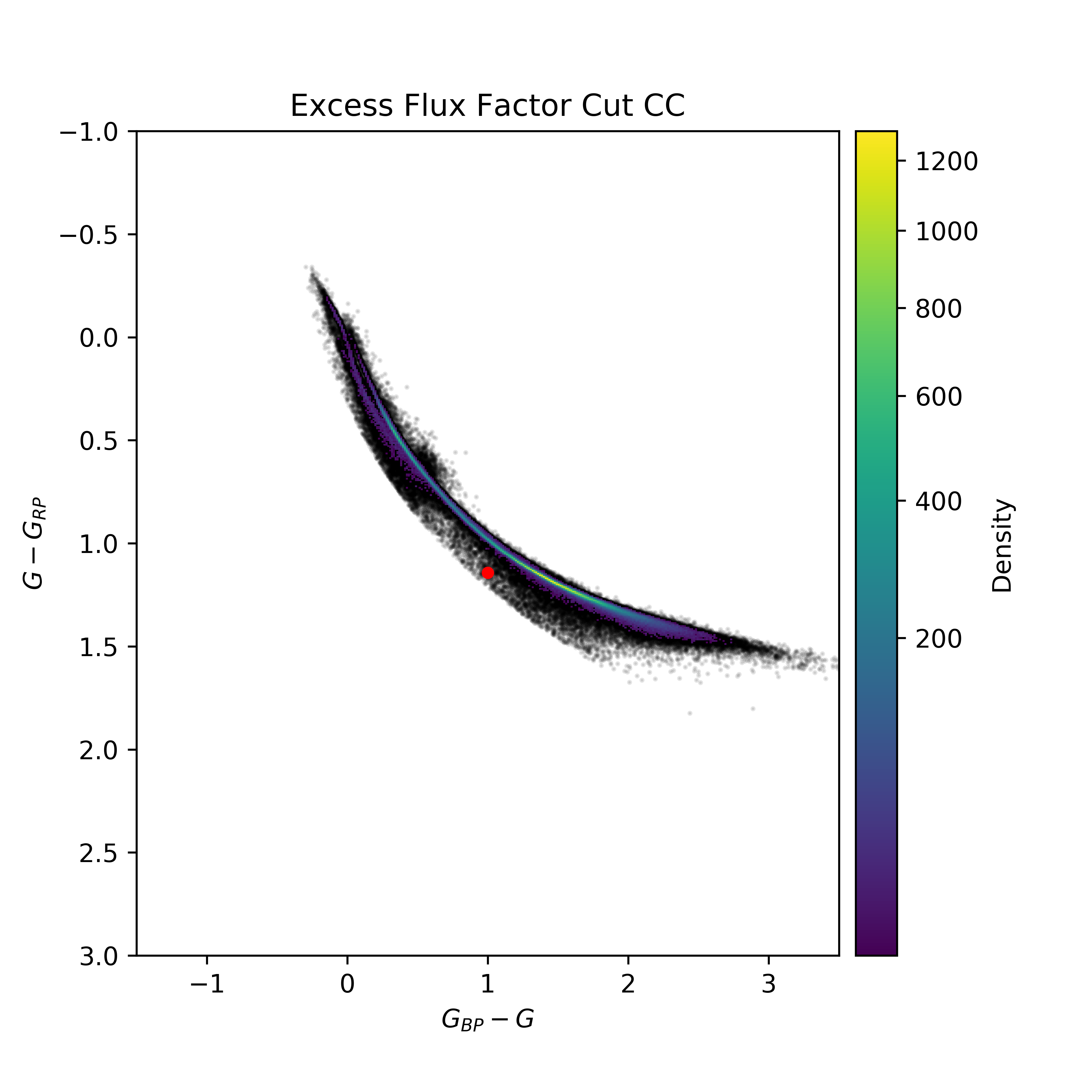} 
\end{tabular}
    \caption{These figures are the HR (top) and CC (bottom) diagrams for the GAIA data sets with finer filter settings. On the left are sources selected with parallax and flux error filters; on the middle are those selected by an additional  astrometric noise cut; and on the right are those that survive an additional excess flux cut. The predicted mirror star region is overlaid on top of HR diagram. 
    }
    \label{f.HRCC}
\end{figure*}

\begin{table*}
    \centering
    \begin{tabular}{|l||l||l|l||l|l||l|l|l|}
    \hline
        \# & Gaia DR2 ID &  Parallax (mas) & d (pc)  & RA (deg) & Dec (deg) & G (mag) & Bp (mag) & Rp (mag) \\
        \hline \hline
1& 2930360704646411904 & 36.03829529 & 27.7
&
112.2433727 & -19.29668099 & 19.519169 & 20.520412 & 18.375805 \\ 
\hline
2
& 4262452645550943232 & 46.46870759 & 21.5
&
285.8209665 & -1.317245813 & 19.78358 & 19.94321 & 17.939726 \\ 
\hline
3
& 4506759318878397568 & 52.51774983 & 19.0
&
285.1298872 & 14.60245224 & 20.026243 & 20.152224 & 17.935917 \\ 
\hline \hline
4
& 4061135605649798400 & 34.8478479 & 28.7
&
262.984546 & -28.29006072 & 19.681303 & 19.908426 & 17.723442 \\ 
\hline
5
& 4064795192593397760 & 42.79213418 & 23.4
&
272.2102654 & -26.22840704 & 19.602968 & 19.313232 & 17.277338 \\ 
\hline
6
& 4064883084836957568 & 38.80994358 & 25.8
&
271.6974055 & -25.74928569 & 19.377226 & 19.568659 & 17.364256 \\ 
\hline
7
& 4077104186615316992 & 249.9741252 & 4.00
&
277.6285842 & -24.70477356 & 20.593948 & 20.919155 & 18.541513 \\ 
\hline
8
& 4111242446355091200 & 26.786845 & 37.3
&
261.3680772 & -24.37027144 & 20.256802 & 20.38141 & 18.364897 \\ 
\hline
9
& 4118771111750851328 & 53.17499257 & 18.8
&
267.41034 & -20.84589839 & 19.128746 & 19.6124 & 17.486 \\ 
\hline
10
& 4118775406720100096 & 55.56323751 & 18.0
&
267.2904763 & -20.76006312 & 19.296127 & 19.808765 & 17.800262 \\ 
\hline
11
& 4120764045300074240 & 170.5518763 & 5.86
&
267.0414042 & -17.44704924 & 19.303873 & 19.32533 & 17.665956 \\ 
\hline
12
& 4314251119361223424 & 24.18581263 & 41.3
&
285.9615572 & 13.23008673 & 20.716675 & 21.419027 & 19.536842 \\ 
\hline
13
& 4514058083198280064 & 48.21504262 & 20.7
&
285.1280652 & 16.94445543 & 20.16315 & 20.768324 & 18.649843 \\ 
\hline
14
& 5928872365503657728 & 29.60234717 & 33.8
&
247.9725741 & -55.47001062 & 20.517174 & 20.784586 & 18.449768 
\\
\hline
    \end{tabular}
    \caption{This table contains the Gaia designation, parallax, right ascension, declination, g mag, bp mag, and rp mag for each of the mirror star candidates. The first candidate (label in bold) is the one we could further examine using other stellar databases. 
    }
    \label{t.mscandidates}
\end{table*}

\subsection{Identifying Mirror Star Candidates}

Mirror stars with optically thin SM nuggets (``optically thin mirror stars'' for short) live in a region of the HR diagram that is very distinct from the areas where regular main sequence stars and white dwarfs live. Furthermore, the visible/IR spectrum of these mirror stars retains the shape of a typical bremsstrahlung spectrum, which is very flat at low frequencies and should, in principle, be distinguishable from regular stars that more closely resemble blackbodies. While the more complete calculation of optically thin mirror stars is likely to add features like line emissions, the fact that an optically thin mirror star is distinguishable from a black body or main sequence stellar spectrum is highly robust.
Optically thin mirror stars are therefore a natural first target, and we now describe a search for these objects in the Gaia data release 2~\cite{2018A&A...616A...1G}. 
A search for optically thick mirror stars requires more careful treatment of the ``white dwarf background'', and is left to future investigation.

We define our optically thin search region as the red box in \fref{f.MSdensity}: $1.61 \leq G_{BP} - G_{RP} < 2.56$ and absolute magnitude $G > 17$. This catches most of the mirror stars in our random scan while staying away from the known locations of main sequence stars and white dwarfs. Since we are searching for very dim objects with relatively large parallax, we restrict our search to objects with distances smaller than 100\,pc, for which distance uncertainties are small.

Following \cite{Lindegren:2018cgr} we then apply some basic quality cuts to the Gaia data, requiring 
\texttt{parallax\_over\_error},  
\texttt{phot\_bp\_mean\_flux\_over\_error} and \texttt{phot\_rp\_mean\_flux\_over\_error}  
all to be greater than 10, i.e., parallax and the two (BP, RP) colour fluxes must be determined with better than 10\% precision. The resulting HR and colour-colour (CC) diagrams are shown in \fref{f.HRCC} (left), and is still contain an unacceptably large contamination of sources intermediate between the main sequence and the white dwarf tract, given that dust extinction within 100\,pc is small. For this reason we apply the \texttt{astrometric\_excess\_noise} $<1$ cut.  The suspect population is significantly reduced, presumably because many binary and crowded or confused sources are rejected \cite{2020arXiv200907277G}. 
The resulting HR and CC diagrams are shown in \fref{f.HRCC} (middle).  

After performing this low astrometric noise cut, our optically thin mirror star signal region contains 14 mirror star candidates, which we list in \tref{t.mscandidates}.  All of them would require further analysis analysis to determine whether they merit follow-up spectral or X-ray observations.
A reason for caution is that all of them have relative magnitudes near Gaia's detection threshold of 20, meaning they represent sources about as dim as Gaia can reliably see.

We now apply some 
further consistency checks on these candidates using Gaia data as well as existing data from additional stellar catalogues. As we will see, only one of them passes the strictest Gaia consistency check, although two others remain candidates of interest.

\begin{figure*}
    \includegraphics[scale=0.45]{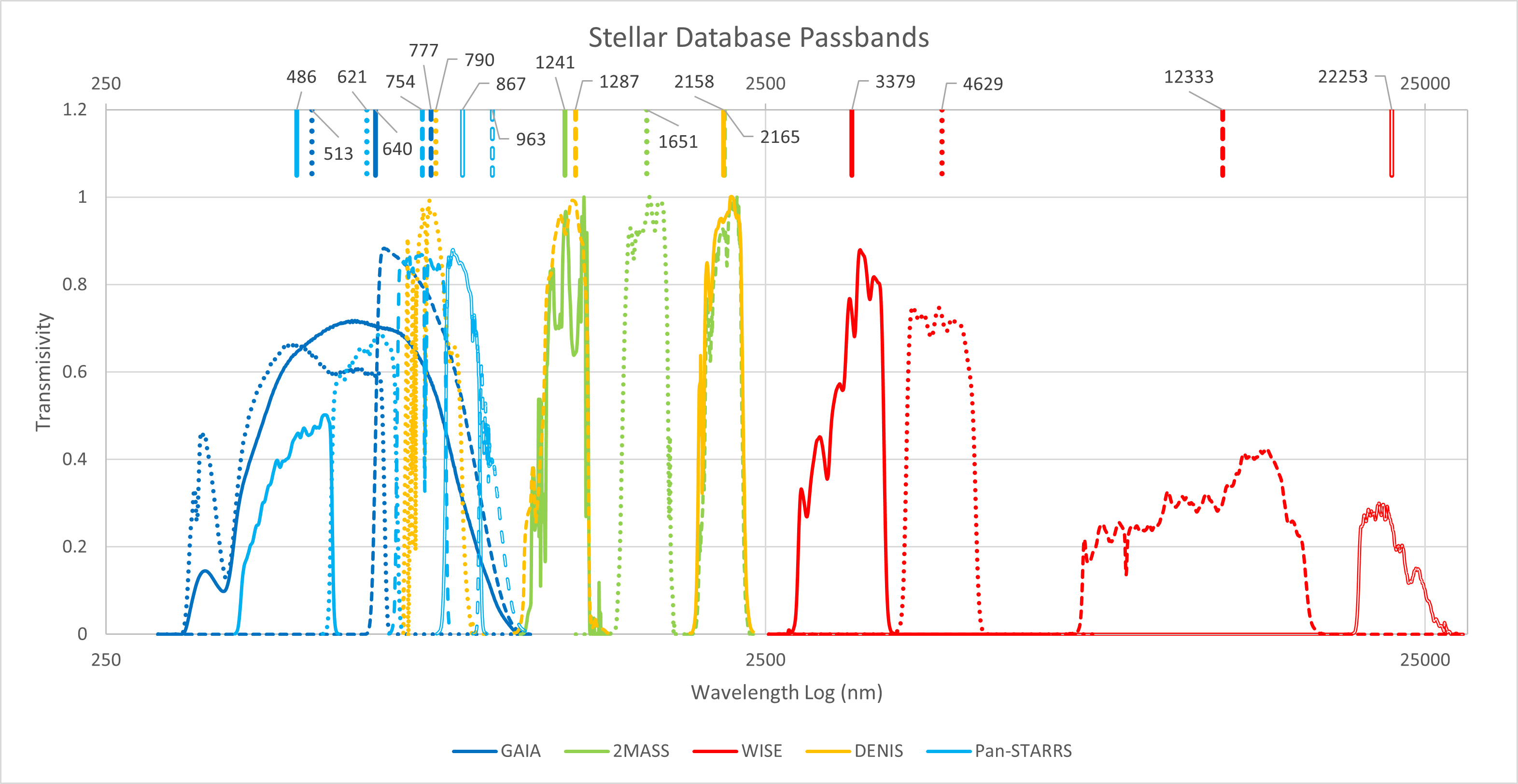}
    \caption{This plots shows all the passband functions used in our analysis, each normalized to unity. Colour corresponds to different databases, while dashing distinguishes different passbands. The vertical lines at the top indicate the transmisivity-weighted  average of each passband, which are used to assign a single frequency value for the flux of each passband in \fref{f.spectrum}.
    }
    \label{f.passbands}
\end{figure*}

\begin{figure*}
\begin{tabular}{cc}
    \includegraphics[width = 0.9 \textwidth]{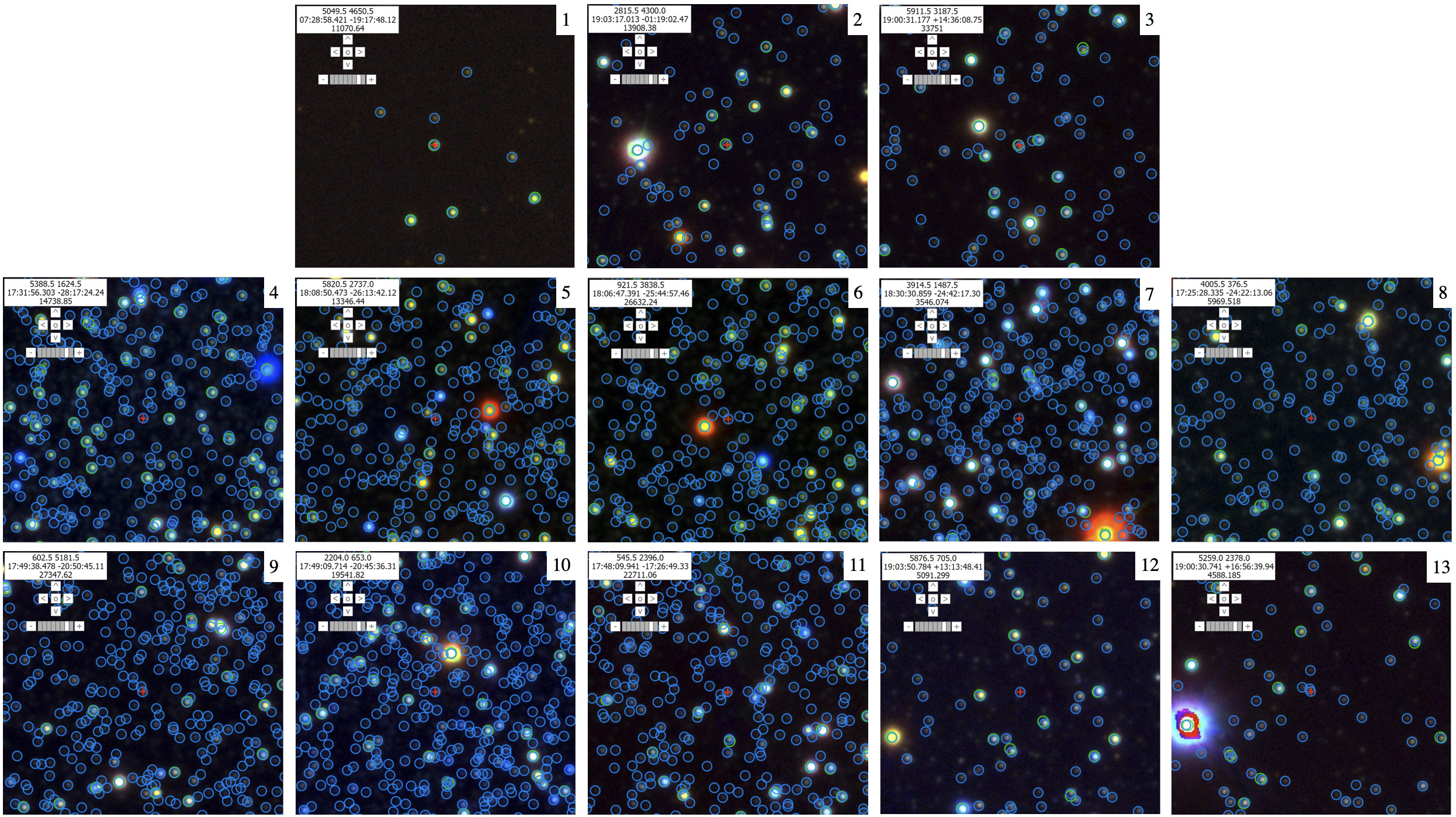} &
    \end{tabular}
    \caption{ Pan-STARRS sky images of mirror star candidates 1 - 13, each marked with a red cross. Blue and green circles indicate the known position of Gaia and 2MASS catalogue objects, respectively. Only the first three candidates  were detected by the other surveys. Candidate 14 is not in a region of the sky imaged by Pan-STARRS. Each of these images is $\sim 1$~arcmin in height and width. 
    }
    \label{f.primecandidate}
\end{figure*}

\begin{table}[]
    \centering
    \begin{tabular}{|l|l||l|l|l|}
    \hline
    $\overline \lambda$ (nm) & Passband & $M_1$ & $M_2$ & $M_3$ \\
    \hline \hline
    486.645 & PanS g &  &  & \\
    \hline
    513.111 & Gaia BP & 20.520 & 19.943 & 20.152 \\
    \hline
    621.463 & PanS r & 21.288 &  & 22.072 \\
    \hline
    640.622 & Gaia G & 19.519 & 19.784 & 20.026 \\
    \hline
    754.457 & PanS i & 20.413 & 21.870 & \\
    \hline
    777.732 & Gaia RP & 18.376 & 17.940 & 17.936 \\
    \hline
    790.727 & Denis I & 18.478 & 17.114 & \\
    \hline
    867.950 & PanS z & 19.920 & 21.109 & 20.332 \\
    \hline
    963.331 & PanS y & 19.549 &  & \\
    \hline
    1241.052 & 2mas J & 16.463 & 16.542 & 16.162 \\
    \hline
    1287.974 & Denis J &  &  & \\
    \hline
    1651.366 & 2mas H & 15.985 & 16.06 & 15.307 \\
    \hline
    2158.701 & Denis K &  &  & \\
    \hline
    2165.632 & 2mas Ks & 15.003 & 16.132 & 14.69 \\
    \hline
    3379.194 & Wise W1 & 15.761 & 12.468 & 10.987 \\
    \hline
    4629.297 & Wise W2 & 15.723 & 12.646 & 11.101 \\
    \hline
    12333.758 & Wise W3 & 12.49 & 11.993 & 10.927 \\
    \hline
    22253.236 & Wise W4 & 8.84 & 8.651 & 8.714 \\
    \hline
    \end{tabular}
    \caption{Absolute magnitudes $M_{1,2,3}$ of our three mirror star candidates in the passbands of the surveys we used. Blank fields indicate no data available.
    }
    \label{tab:my_label}
\end{table}

Detailed spectral data are not commonly available for an arbitrary star in our galaxy, although surveys such as SDSS and LAMOST provide increasing coverage \cite{gentile2019gaia,Kong21_WDs_in_LAMOST}.
However, various stellar catalogues contain absolute magnitude information over a range of wavelengths as defined by their passband functions. If we can cross-match the mirror star candidates in \tref{t.mscandidates} with objects detected in other surveys, we can assemble the magnitude information from all the surveys' passbands into a crude spectral measurement of the object over a range of wavelengths.
The catalogues we use, as well as their approximate wavelength ranges and limiting magnitudes, where available, are the following. 
    \begin{itemize}
        \item Gaia~\cite{2016A&A...595A...1G}: 300-1,100nm $(M_G<18, M_{BP}<18,M_{RP}<18)$
        \item Pan-STARRS~\cite{chambers2019panstarrs1}: 400-1100nm $(M_g<23.3, M_r<23.2, M_i<23.1, z<22.3, y<21.3)$
        \item DENIS~\cite{1993ASPC...52...21D}: 700-2500nm $(M_I<15.5, M_J<14.5, M_K<13)$
        \item ALLWISE~\cite{Kirkpatric14:allwise}: 1000-2500nm $(M_{W1}<17.1, M_{W2}<15.7, M_{W3}<11.5, M_{W4}<7.7)$ 
        \item 2MASS~\cite{2006AJ....131.1163S}: 2500-30000nm $(M_J<15.8, M_H<15.1, M_{Ks}<14.3)$
    \end{itemize}
We use the GATOR utility \cite{GATOR} to cross reference Gaia candidates with the other databases.
The passbands from all these surveys are shown in \fref{f.passbands}. We also indicate the transmissivity-weighted average 
of each passband at the top of the plot, which we use to assign each passband's flux measurement a single wavelength value when assembling a crude  spectral energy distribution for each mirror star candidate.

In \fref{f.primecandidate} we show Pan-STARRS sky images for the first 13 mirror star candidates.
Most of the candidates in \tref{t.mscandidates} were not detected in the DENIS, ALLWISE or 2MASS catalogues. This is not surprising, as Gaia is significantly more sensitive. Only the first three candidates show up in the majority of the other catalogues.

Some further insight might be gained by applying a final quality cut on the Gaia data. 
Both blackbody and bremsstrahlung spectra of various temperatures should lie along the high-density line in the CC diagrams of \fref{f.HRCC}. This can be easily understood, since the RP and BP Gaia passbands measure complementary portions of the wavelength range of the G passband, and their independent fluxes should approximmately sum to the G flux.  A strong deviation from the narrow region of consistency in the CC diagram could indicate a measurement error of one or all of the passband magnitudes, or contamination of the measurement by other sources close by in the sky.
In fact, a visual inspection of the Pan-STARRS images in \fref{f.primecandidate} seems to suggest that all the candidates apart from \#1 and \#2 might suffer from source confusion, either due to identified nearby sources or due to unidentified background sources. 

The Gaia collaboration~\cite{Lindegren:2018cgr} suggests an excess flux factor cut to eliminate such objects with potential errors in their colour measurement:
$1+0.015 \times (\texttt{bp\_rp})^2< \texttt{phot\_bp\_rp\_excess\_factor} <1.3+0.06 \times(\texttt{bp\_rp})^2$.
After applying this cut, we are left with the HR and CC diagrams in \fref{f.HRCC}, and only our first mirror star candidate in the optically thin signal region. 
This makes candidate \#1 our primary mirror star candidate for additional study, though we will also examine \#2 and \#3 since data in other catalogues is available for them as well. The other candidates could still in principle be mirror stars, and in a realistic analysis we should conduct follow-up analyses to see if they are present in any other catalogues. However, for the purposes of demonstrating our search pipeline in this toy analysis it is sufficient to focus on the first three and most high-quality candidates.

\subsection{Examination of candidates \#1, \#2 and \#3}

For any mirror star candidate that is captured by multiple surveys, we  can assemble a pseudo-spectrum of magnitude measurements $(\overline \lambda_X, M_X)$ in the different passbands $X$, where $\overline \lambda_X$ is the 
average wavelength of the passband with transmissivity  $P_X(\lambda)$,
\begin{equation}
\label{e.avglambda}
\overline \lambda_X = \frac{\int P_X(\lambda)\lambda d\lambda}{\int P_X(\lambda) d\lambda},
\end{equation}
see \fref{f.passbands}.
The magnitude $M_X$ in each passband is related to the  flux density $F(\lambda)$ (taken to be the flux at 10pc) as follows \cite{2010A&A...523A..48J, 2018MNRAS.479L.102C}:
\begin{equation}
\label{e.passbandmag}
    M_X = -2.5 \log_{10} \left( \frac{A}{10^9 h c} \int_{\lambda_{min}}^{\lambda_{max}} d\lambda \, F(\lambda) P_X(\lambda) \lambda \right) + ZP_{X},
\end{equation}
where $\lambda$ is in nm, $ZP_X$ is the zero point\footnote{In cases where the zero point for a given passband was not easily available, we derived it using the known flux from Vega~\cite{2018ascl.soft11001S} and the fact that the observed magnitude of Vega is zero in any band by definition. For passbands where the zero point was easily available, this agreed perfectly with the official values.} of band $X$, and $A$ is the telescope pupil area for each survey.

\begin{figure}
    \centering
    \includegraphics[scale=0.32]{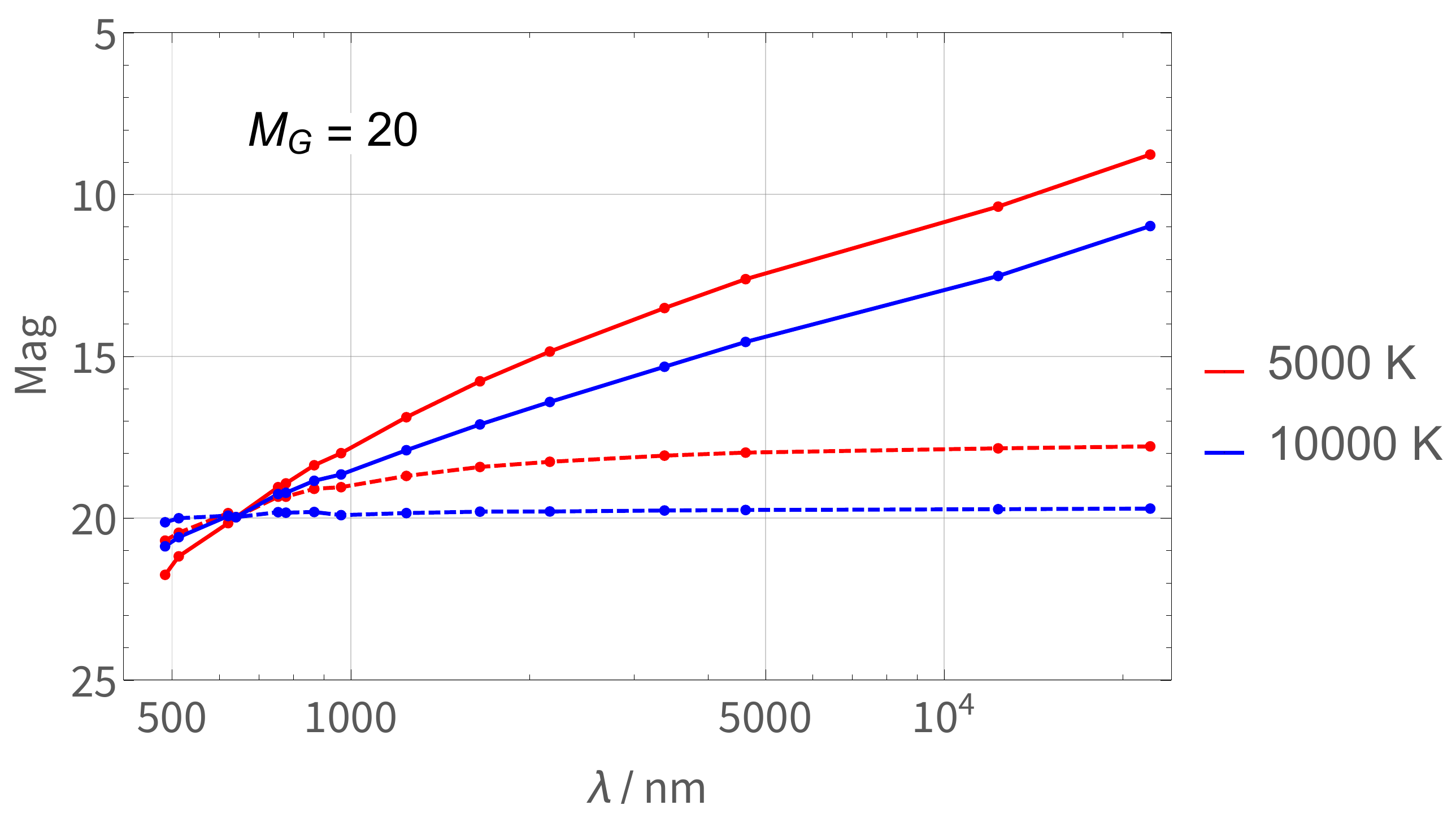}
    \caption{
    Solid: Predicted bremsstrahlung-like pseudo-spectrum of a mirror star with an optically thin SM nugget at a temperature of 5000K (red) and 10000K (blue) in the simplified signal calculation of~\cite{Curtin:2019lhm}. Dashed: blackbody at same temperature for comparison. 
    Each dot $(\overline \lambda_X, M_X)$ corresponds to a survey passband X, see Eqns.~(\ref{e.avglambda}) and~(\ref{e.passbandmag}). 
    All curves are normalized to an apparent magnitude of 20 in Gaia's $G$ passband.
    }
    \label{f.spectrumpredictions}
\end{figure}

Since optically thin mirror stars have a bremsstrahlung-like spectrum in our simplified signal calculation, their pseudo-spectrum could be distinguishable from the expected blackbody-like spectrum of regular stars. We demonstrate this in \fref{f.spectrumpredictions}, where we plot the predicted pseudo-spectrum for optically thin mirror stars and blackbodies at 5000K and 10000K. 
The difference between the mirror-star prediction and the blackbody is much greater than the variation due to the expected range of possible temperatures, with optically thin mirror stars being comparatively much brighter at low frequencies. (The increase at high wavelengths for the ``flat'' bremsstrahlung spectrum is due to the $\lambda$ factor in the integral of \eref{e.passbandmag}.)

\begin{figure}
\begin{tabular}{c}
        \includegraphics[height=5cm]{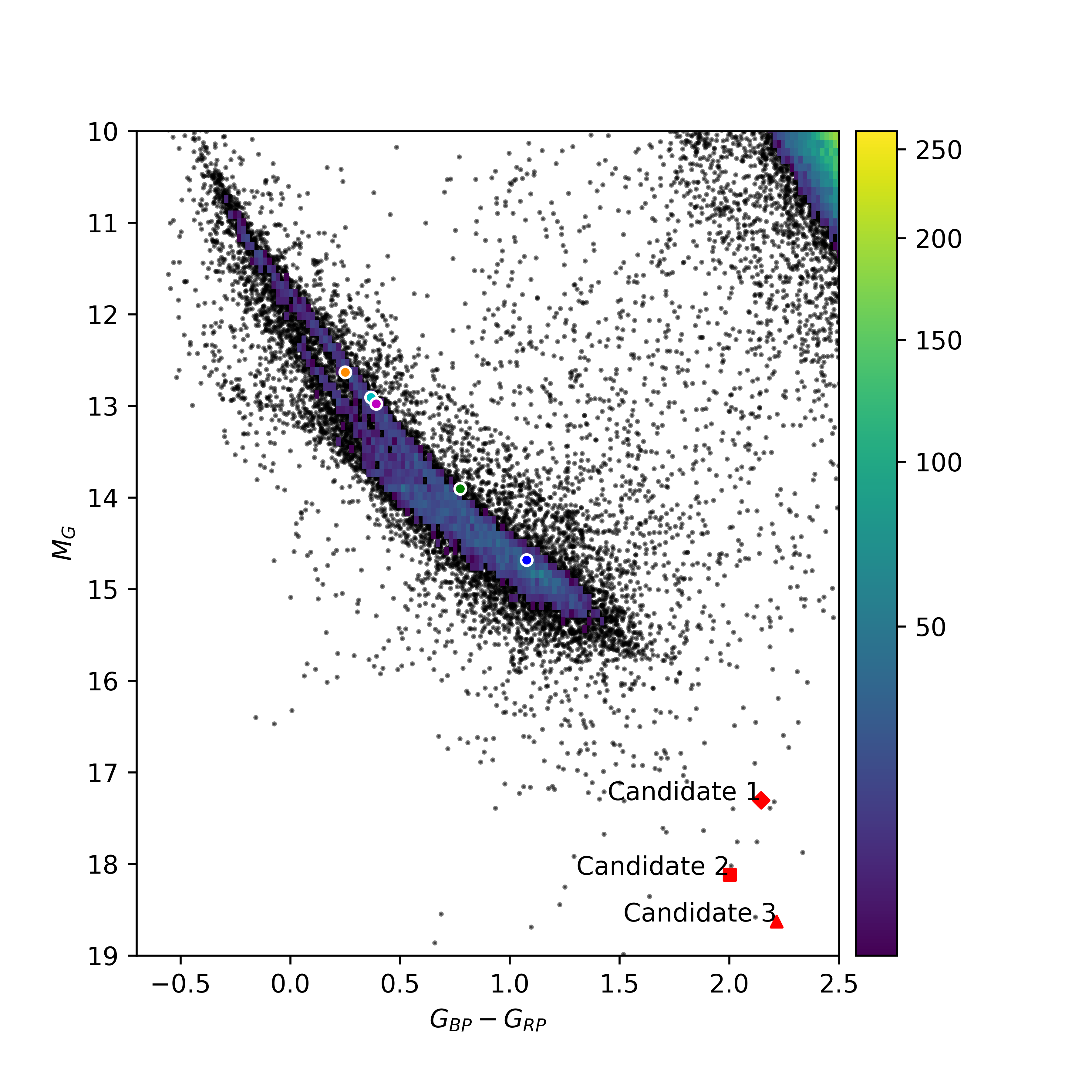} \\
        \includegraphics[height=4cm]{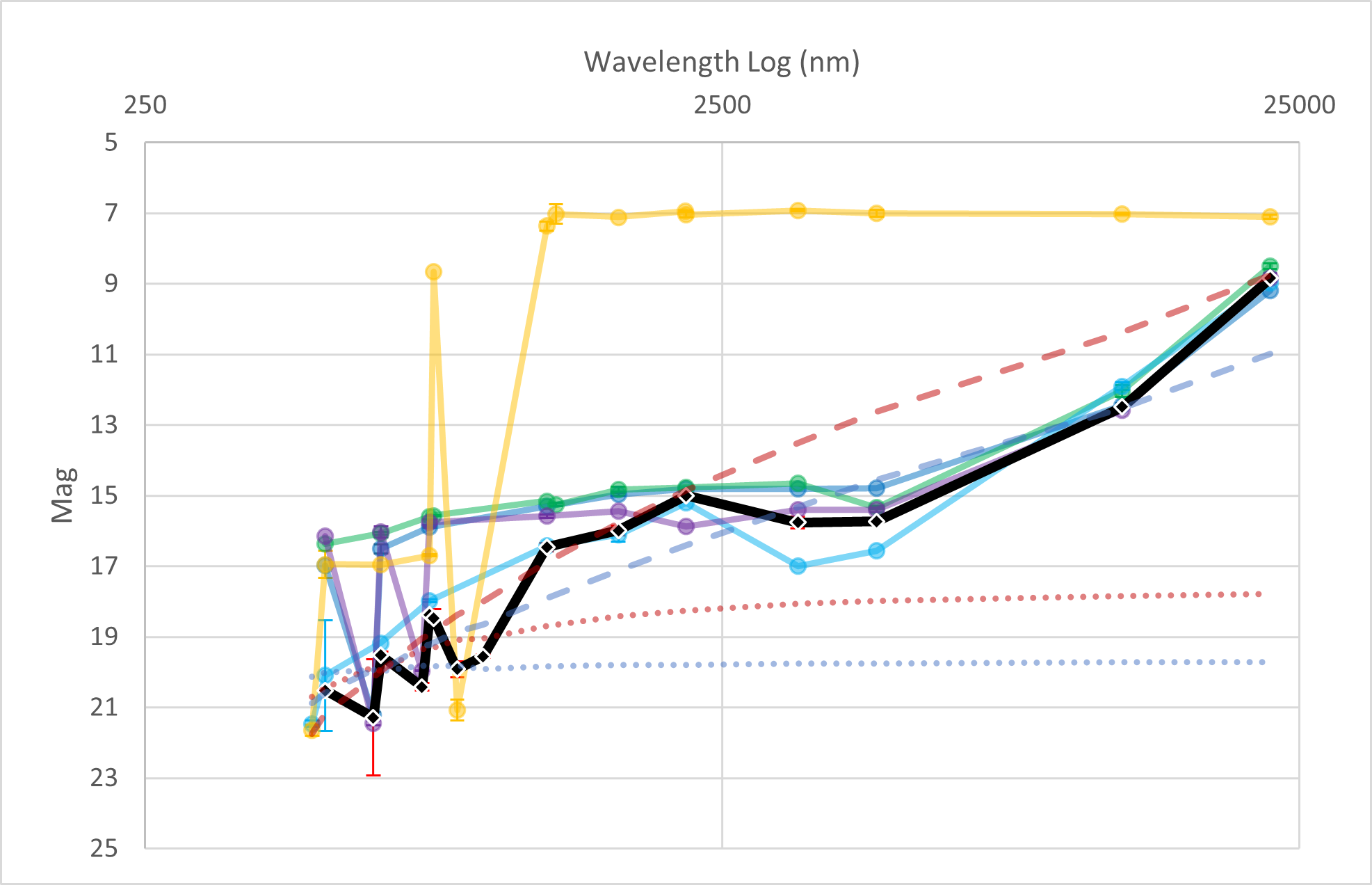} \\
        \includegraphics[height=4cm]{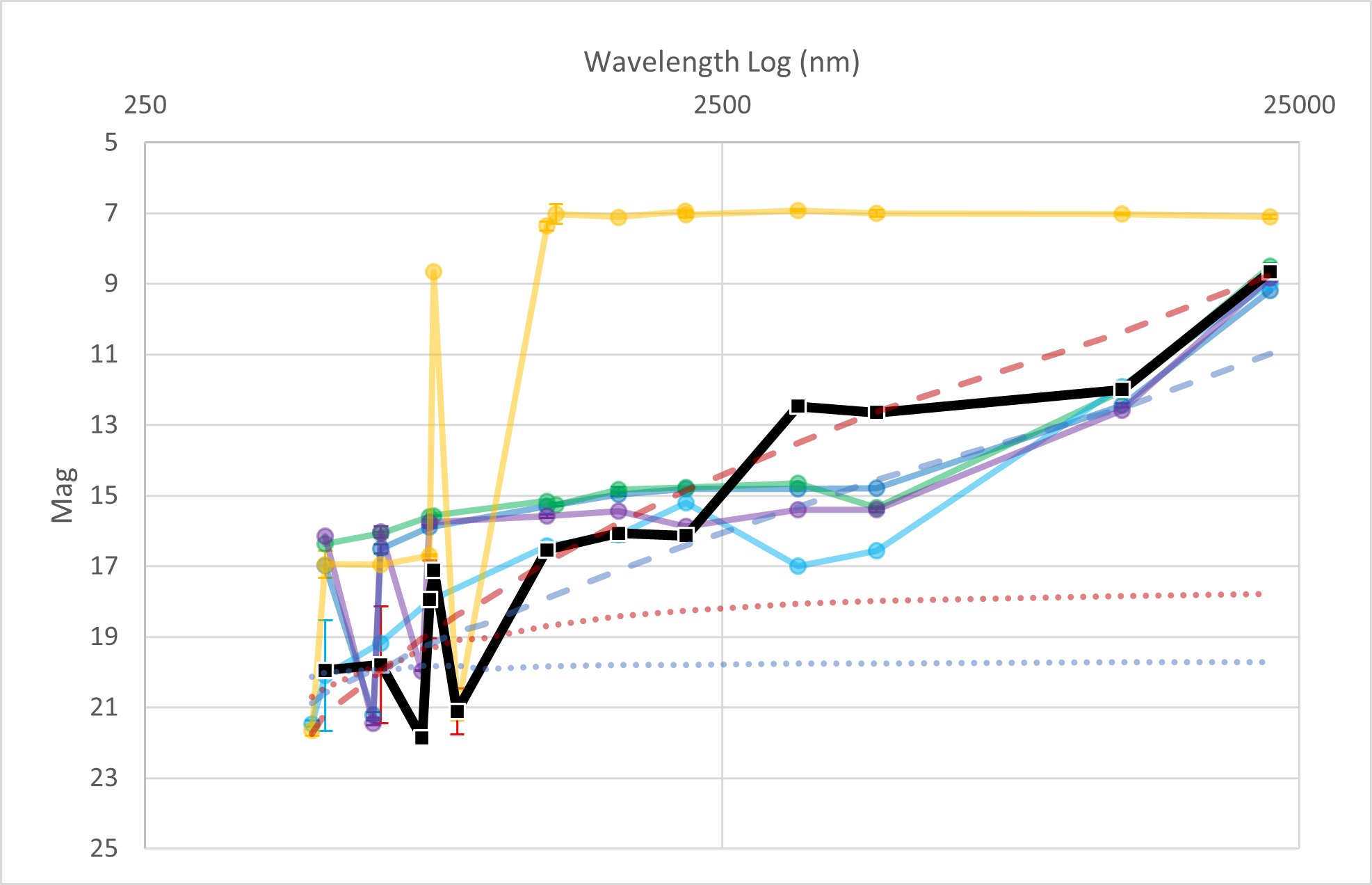} \\
        \includegraphics[height=4cm]{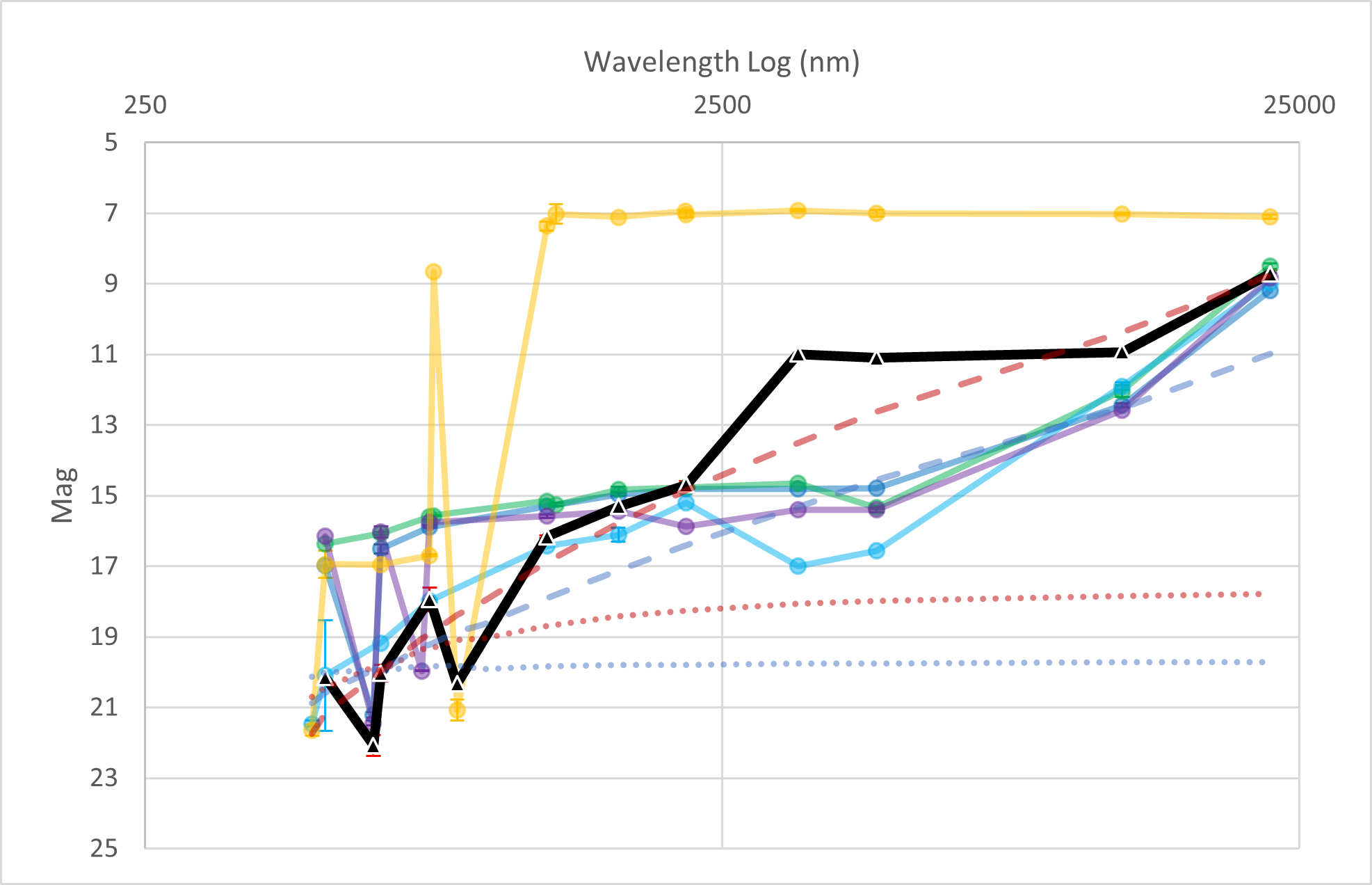} \\
        \includegraphics[height=2.8cm]{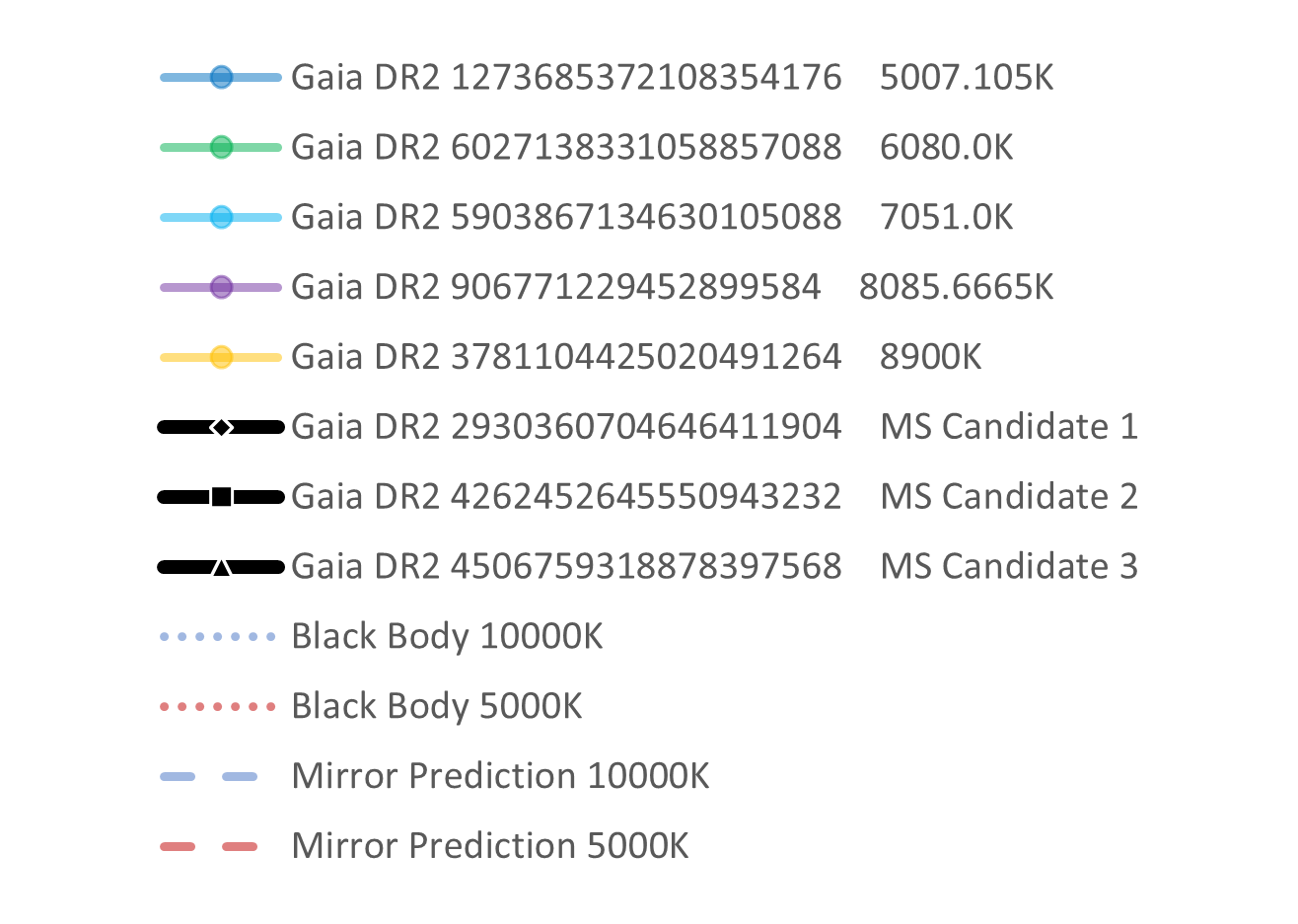}
    \end{tabular}
    \caption{
    HR-diagram shows white dwarf region and mirror star candidates
    \#1, \#2, \#3. 
    Also marked are 5 randomly selected comparison white dwarfs with temperatures from 5000K - 9000K. 
    Plots show pseudo-spectra of mirror star candidates (black solid)
    and comparison white dwarfs (coloured solid), 
    using magnitude measurements in different stellar survey passbands.  The dashed/dotted opaque curves are the mirror star / blackbody theoretical expectations from \fref{f.spectrumpredictions} for comparison.
    }
    \label{f.spectrum}
\end{figure}

Pseudo spectra for mirror star candidates \#1, \#2, and \#3 are shown  in \fref{f.spectrum} (red diamonds, squares and triangles).
These pseudo-spectra display the increase at high wavelengths that might naively be suggestive of  bremsstrahlung-like emissions. 
However, it is not clear if the departure from the expected blackbody shape could be explained by more benign astrophysics.

For guidance from the data, we randomly select 5 objects from the white dwarf region of the Gaia HR diagram with observed temperatures in the range of 5000$K$ -- 9000$K$ (marked in HR diagram cutout in \fref{f.spectrum}), and assemble their pseudo-spectra from the stellar catalogues' data in the same way as for the mirror star candidates. Their pseudo-spectra are shown as the coloured dots in \fref{f.spectrum} (right), and look  remarkably similar to our mirror star candidates.  
Obviously, 
distinguishing mirror stars' bremsstrahlung spectra from white dwarfs using
multi-band colours
is not straightforward, since benign astrophysical effects apparently give rise to a similar luminosity increase at long wavelengths as predicted for optically thin mirror stars. 
Reliably separating mirror stars from white dwarfs in this way would likely require a more detailed analysis, but may be helped by additional  emission processes of the SM nugget that we have not included in our simplified signal calculation.

\begin{figure}
        \includegraphics[width=7cm]{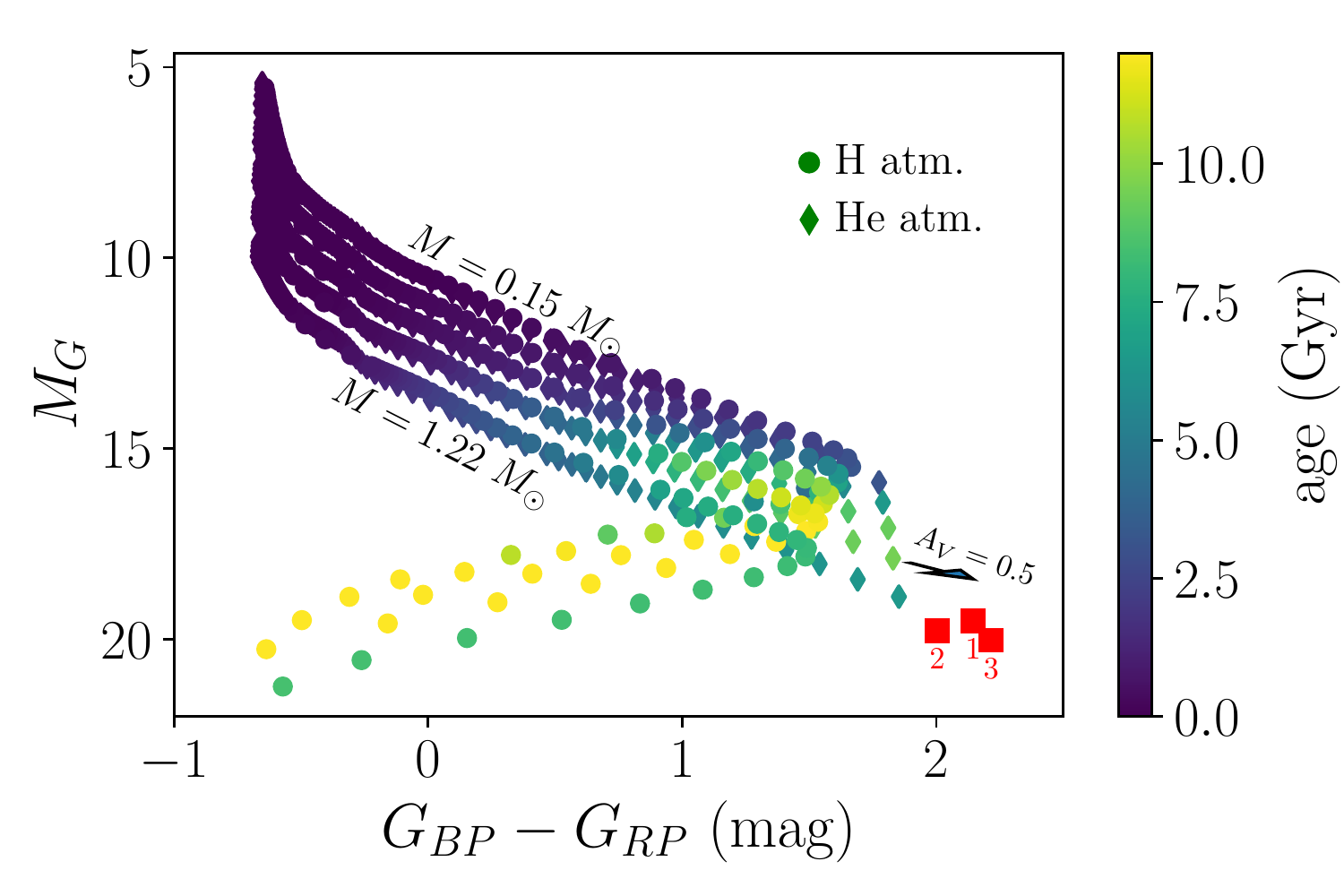}
    \caption{
    {\it Gaia} colour-magnitude diagram comparing Candidates \#1-\#3 to model    cooling sequences of hydrogen-atmosphere and helium-atmosphere white dwarfs. The models plotted derive from \cite{holberg2006calibration}, \cite{blouin2018new}, \cite{bedard2020spectral}, and \cite{tremblay2011improved}, combined and disseminated in \cite{BergeronSite}. Also shown is a reddening vector, corresponding to 0.5 visual magnitudes of dust extinction as estimated by  \cite{gentile2019gaia}.  
    }
    \label{f.WDs}
\end{figure}

Given the comparison, it seems likely that these mirror star candidates are simply particularly dim white dwarfs.  In favour of this hypothesis, we note that their location in the colour-magnitude diagram would be very close to the region populated by old and massive helium-atmosphere white dwarfs \cite{Torres21_GaiaWDs} with a small amount of dust extinction. \fref{f.WDs} demonstrates this possibility. 
However, their emission is also not inconsistent with them being optically thin mirror stars in our emission model, and they are significantly removed from the highly populated white dwarf region of the HR diagram. 
Follow-up optical spectroscopy would be ideal to settle this issue.

\section{Constraining Mirror Stars}
\label{s.MSconstraints}

Even if Gaia detects no mirror stars, it will still yield important \emph{constraints} on the mirror stars density.
Fig.~\ref{fig:HR_density} shows how each mirror star corresponds to a dot on the HR diagram, and that non-detection can be translated into a 95\% confidence level upper bound (4 expected but 0 observed signals) on the local number density of such mirror stars.
In this section, we show how to translate a non-detection by Gaia into constraints on mirror star \emph{properties}, like mass and core temperature. In principle, this might be connected to fundamental parameters of the hidden sector Lagrangian in the future.

Note that our simplified model of mirror star emission may not be robust with respect to the precise colour of the SM nuggets, but it is likely to give a good estimate of its bolometric luminosity. We therefore derive these constraint projections without explicit reference to any particular signal region in the HR diagram, since this may change with a more sophisticated calculation. Rather, we simply assume that optically thin mirror stars are \emph{distinctive} and that a signal region can be defined within which non-mirror-stars can be eliminated. This is justified given how different the emission mechanisms of optically thin nuggets are compared to normal stars. The constraints on optically thin mirror stars are therefore assumed to derive from a background-free search without any positive detection. For optically thick mirror stars we consider two possibilities, one optimistic case where they can be distinguished from regular stars, resulting in a zero-background-search, and one pessimistic case where all white dwarfs count as backgrounds. Our results will therefore give a good idea of Gaia's sensitivity.

Even this effective approach, which does not start from hidden sector microphysics, is extremely challenging in generality. Likelihood of observation depends not just on the properties of a given mirror star, but also on their mass and age distribution in our stellar neighborhood. We therefore start with a non-physical warm-up by assuming that mirror star properties and distributions are identical to SM stars. This teaches us which properties of mirror star distributions most affect constraints, allowing us to use only the properties of ``mirror red dwarfs'' to obtain simplified constraints on the mirror star density as a whole.

\subsection{Warm-up: Constraining a symmetric mirror sector}

As a simple warm-up, we will assume that mirror-stars are perfectly SM-like, and that their mass distribution is given by  the IMF of SM stars given by \cite{Kroupa:2000iv}:
\begin{equation}
    \label{e.IMF}
    P(M_{MS}) \propto \left(\frac{M_{MS}}{M_\odot}\right)^\beta
\end{equation}
where
\begin{equation}
\beta = \left\{ 
    \begin{array}{ccc}
    -0.3 & \mathrm{when} & M_{MS}/M_\odot < 0.08,
    \\
    -1.3 & \mathrm{when} & 0.08 \le M_{MS}/M_\odot < 0.5,
    \\
    -2.3 & \mathrm{when} & 0.5 \le M_{MS}/M_\odot,
    \end{array} \right\}
\end{equation}
and the various piece-wise parts are continuously joined and normalized to a given total mirror star number density.
Even if the mirror sector Lagrangian were an exact copy of the SM, this scenario is not  realistic, since in general the shape of the mirror star mass distribution is influenced by the dynamics of mirror matter in our galaxy, which depends on the fraction of dark matter that is mirror matter. However, for the purpose of illustration we ignore this complication and assume that the mass function has a fixed shape with only the normalization scaling directly with the local mirror star number density $n_{MS}$, and derive constraints on $n_{MS}$ as a function  of $\epsilon \sqrt{\alpha_D/\alpha_{em}}$.

For different stellar masses above $0.14 M_{\odot}$, we use the MESA stellar evolution code~\cite{Paxton_2010,Paxton_2013,Paxton_2015,Paxton:2017eie,Paxton_2019} to obtain the stars' radius, core density, core temperature, and main sequence lifetime. We then generate a large sample of mirror stars following the mass function, with each star's age chosen randomly from its lifetime. 
For a given $\epsilon \sqrt{\alpha_D/\alpha_{em}}$ we can then compute the bolometric magnitude in SM photons for each mirror star, and then compute the total number of stars Gaia is expected to observe as a function of their number density:
\begin{multline}
    N_{obs} = n_{MS} \int_0^{D_{max}} dD \, 4\pi D^2  \\ 
    \sum_i w_i \Theta(G_{max} - m_{app}(m_{abs}, D)),
\end{multline}
where $m_{app}$ is the apparent magnitude as a function of the absolute magnitude $m_{abs}$ and the distance $D$, $G_{max}$ is the maximum magnitude we assume Gaia can see, $D_{max} = 100$ pc is the maximum assumed mirror star distance for reliable detection and parallax, and $w_i$ are the relative weights assigned to each star according to their mass distribution, defined such that $\sum_i w_i = 1$. The sum is over mirror stars above Gaia's detection threshold, since the Heaviside function determines whether each star is visible at distance $D$. To obtain our constraints we then find the value of $n_{MS}$ corresponding to a number of expected mirror star observations $N_{obs}$ that non-observation at Gaia can exclude, see below. 
$N_{obs}$ is computed separately for optically thin and thick mirror stars in our sample, and we take whichever of the two resulting constraints on $n_{MS}$ is stronger.

\begin{figure}
    \includegraphics[scale=0.55]{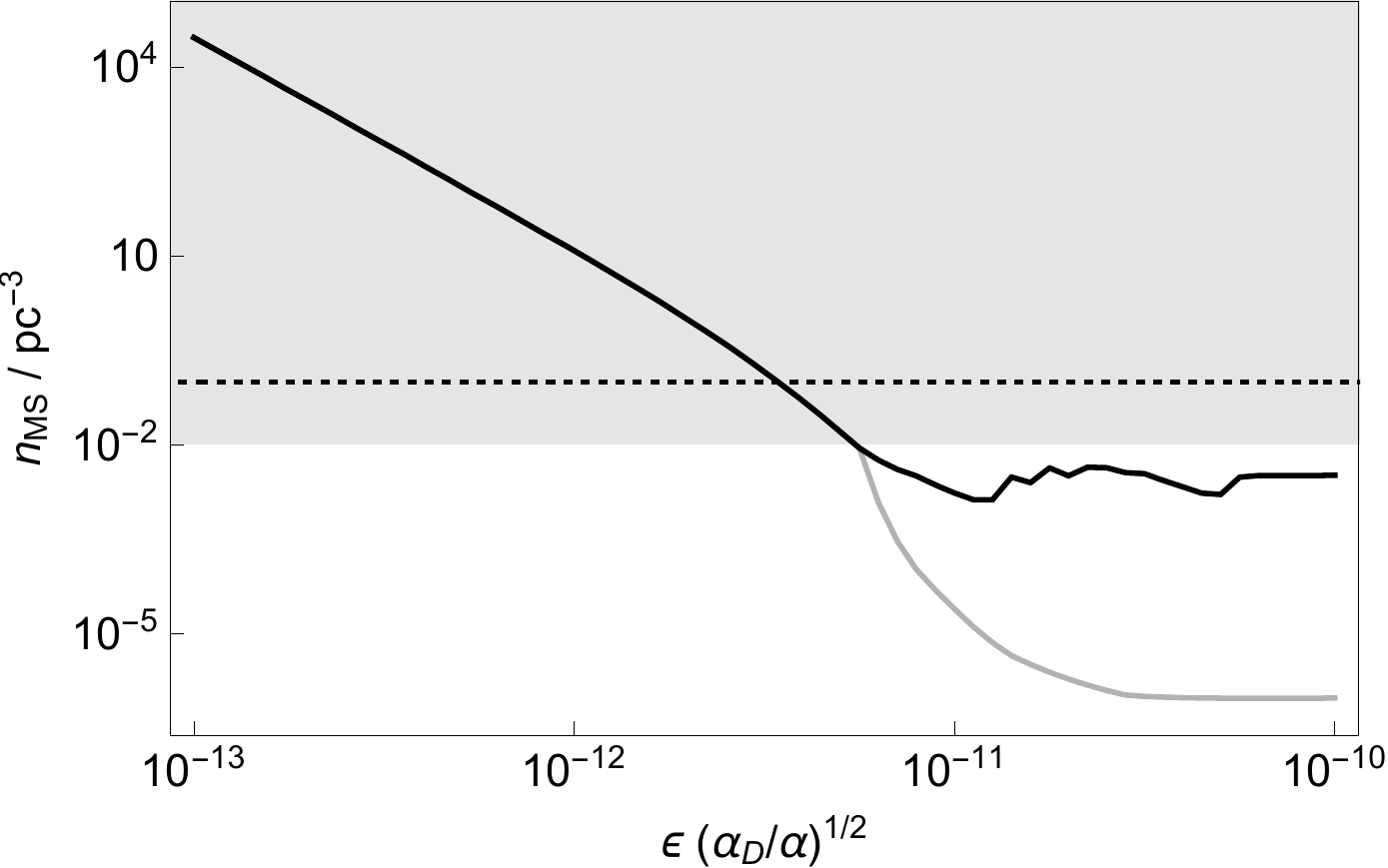}
    \caption{Constraints on mirror star number density as a function of the kinetic mixing parameter and dark QED coupling, assuming SM-like mirror stars with a SM-like mass function.
    For $\epsilon \ll 10^{-11}$, the dominant constraint comes from requiring the expected number of observed optically thin mirror stars to be less than 4. For $\epsilon \gtrsim 10^{-11}$, the limit is dominated by optically thick mirror stars, with the black (gray) curve corresponding to using current (optimistic future) limits on their observed number.  
    In the gray shaded region, the mirror star mass density excluded by Gaia exceeds the local dark matter density.
    The black dotted line is the local number density of regular stars.
    }
    \label{f.SMcopybounds}
\end{figure}

\begin{figure}
    \includegraphics[scale=0.6]{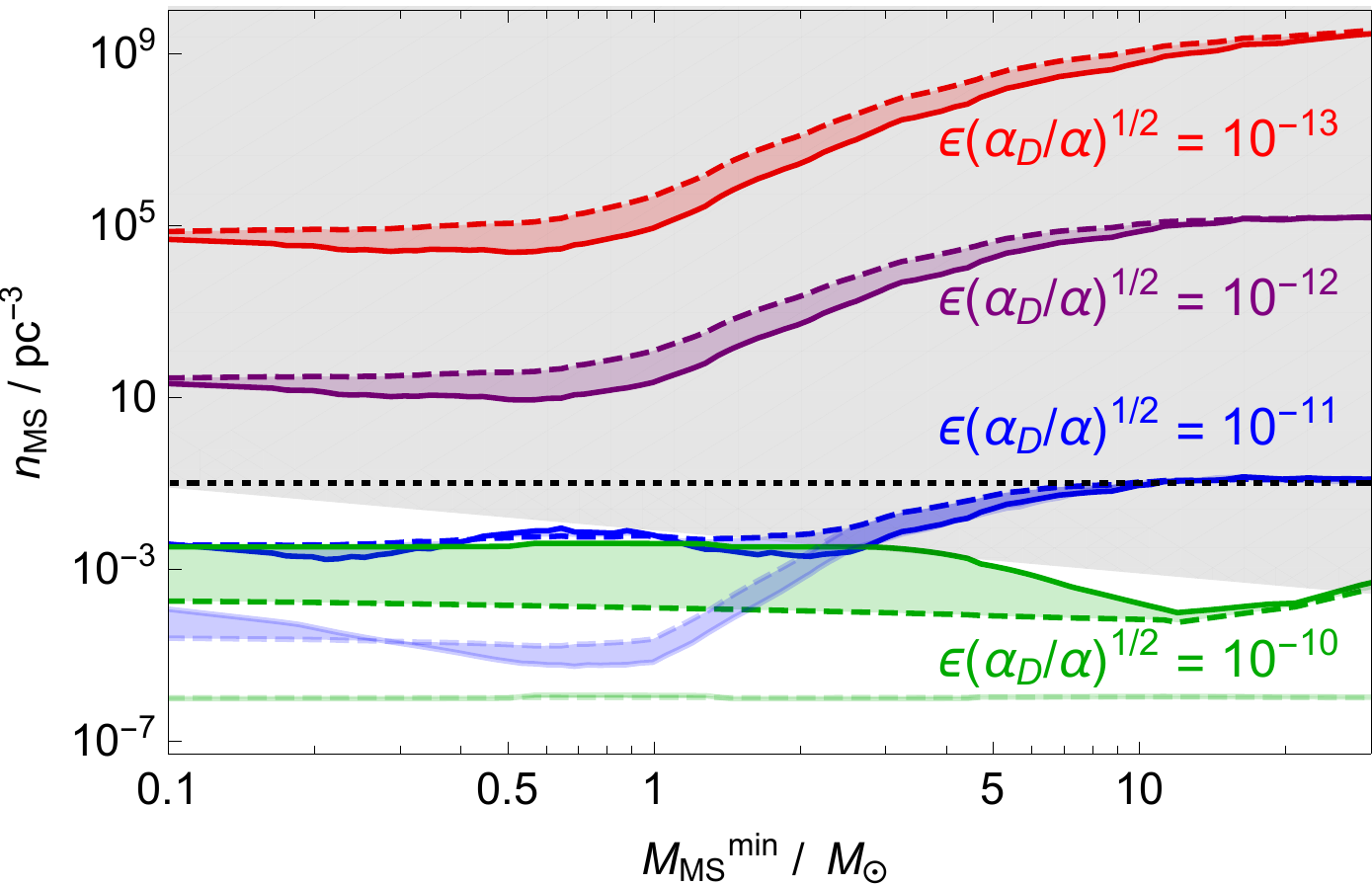}
    \caption{Gaia constraints on SM-like mirror star number density, for $\beta = -1$ mass function with minimum stellar mass $M_{MS}$ (dashed), or delta function mass distribution (solid). The coloured curves correspond to four different values of the kinetic mixing parameter times dark coupling constant. The shading indicates the effect of the unknown mirror star mass function.
    The more (less) opaque curves correspond to the current (optimistic future) constraints derived from the optically thick mirror stars. 
    In the gray region, the mirror star mass density excluded by Gaia exceeds the local dark matter density.
    The black dotted line is the local number density of regular stars.
    }
    \label{f.deltafnbounds}
\end{figure}

Optically thin mirror stars live in a region of the HR diagram populated by very few regular stars, so we set constraints assuming no signal has been observed with zero expected background events. The limit $n_{MS}^\mathrm{thin}$ then corresponds to $N_{obs} =  4$ expected mirror star observations. For optically thick mirror stars, a more sophisticated analysis would be needed to obtain constraints that account for the ``background'' of regular white dwarf stars. 
Such an analysis is beyond our scope, but we can still derive some constraints by requiring that the total number of expected optically thick mirror stars does not \emph{exceed} the number of observed white dwarfs in the Gaia dataset of our search, corresponding to $N_{obs} = N_{WD} \approx 1.36 \times 10^5$ within 100 pc. 
This is extremely pessimistic but represents an actual constraint of our analysis. 
Clearly, a realistic analysis could do much better, but the best it could ever do in principle is perfect discrimination of white dwarfs from mirror stars (e.g. via detailed spectral or X-ray measurements), in which case the search for optically thick mirror stars becomes effectively background free and a projected constraint can be set on $n_{MS}^\mathrm{thick}$ by requiring $N_{obs} = 4$ expected observed events. 
We  refer to these two assumptions for optically thick mirror stars as the current and optimistic future constraints respectively. 
Finally, we show combined constraints on $\epsilon (\alpha_D/\alpha)^{1/2}$ from optically thin and optically thick mirror stars in \fref{f.SMcopybounds} by simply using whichever of the two constraints is stronger and requiring $n_{MS} < \mathrm{min}(n_{MS}^\mathrm{thin}, n_{MS}^\mathrm{thick})$.\footnote{This simple combination is sufficient since the huge majority of mirror stars is either optically thin or optically thick depending for low or high $\epsilon (\alpha_D/\alpha)^{1/2}.$}

In this example, the optically thick mirror star  search would supply the strongest current (optimistic future) constraint for $\epsilon \sqrt{\alpha_D/\alpha_{em}} \gtrsim 3 \times 10^{-11} (5 \times 10^{-12})$, while for smaller kinetic mixings, the strongest constraint derives from the optically thin mirror star search. This is reasonable, since for larger kinetic mixings, the interaction between SM matter and the mirror matter is increased, leading to larger and hotter SM nuggets.
The current limits basically restrict the optically thick mirror star number density to be smaller than the white dwarf number density, which is roughly an order of magnitude below the total stellar number density (dotted horizontal line), while the optimistic assumption for a possible future limit is about three orders of magnitude stronger, demonstrating the sensitivity gain achievable with a dedicated mirror star observational search.

\begin{figure*}
    \centering
    \includegraphics[width=0.8 \textwidth]{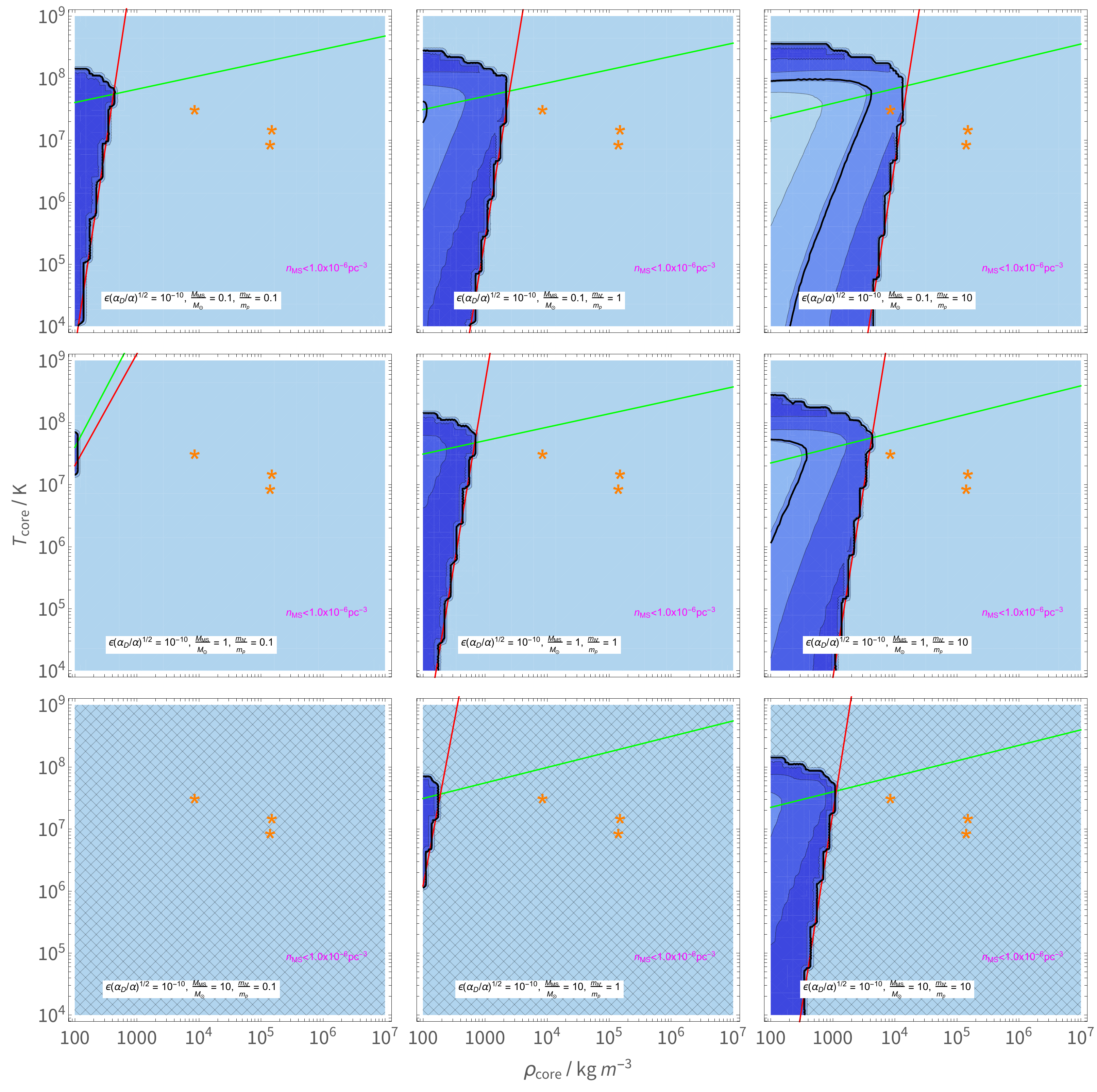}
    \includegraphics[scale=0.662]{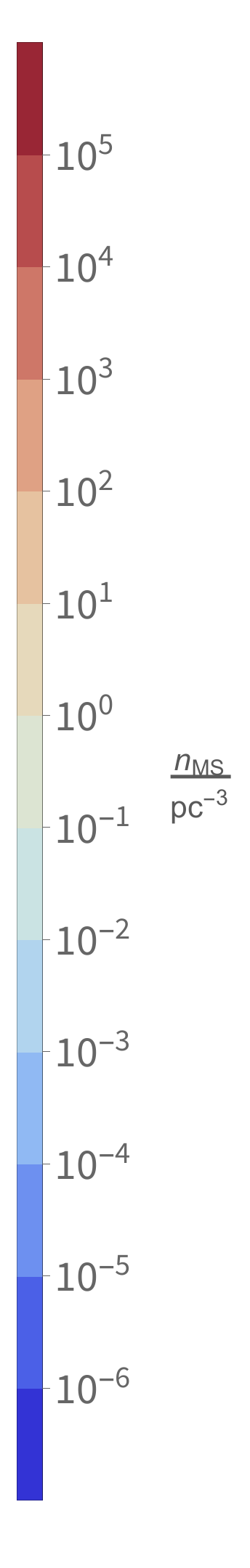}
    \caption{
    Gaia constraints on the local Mirror Star number density $n_{MS}$ as a function of core density and core temperature, for $\epsilon \sqrt{\alpha_D/\alpha_{em}} = 10^{-10}$. From top to bottom, the panels assume that mirror red dwarfs have mass $M_{MS}/M_\odot = 0.1, 1, 10$ and from left to right assume nuclear constituent masses $m_N/m_p =  0.1, 1, 10$. The hatched region indicates where the Gaia constraints are weaker than the requirement that $\rho_{MS}$ is smaller than the locally measured DM mass density $\rho_{DM} = 0.5~\mathrm{GeV}/\mathrm{cm}^{-3}$. 
    To the left (right) of the red line,  the optically thin (thick) region provides
    the strongest constraint. %
    Above the green line, converted X-rays from the mirror star core contribute to heating the SM nugget.
    The solid black contour corresponds to $\rho_{MS} = 0.01 \rho_{DM}$. The three orange asterisks indicate $(\rho_{core}, T_{core})$ of a $0.2, 1$ and $10$ solar mass SM star (in order of increasing core temperature) for reference.}
    \label{fig:grid10}
\end{figure*}

\begin{figure*}
    \centering
    \includegraphics[width=0.8 \textwidth]{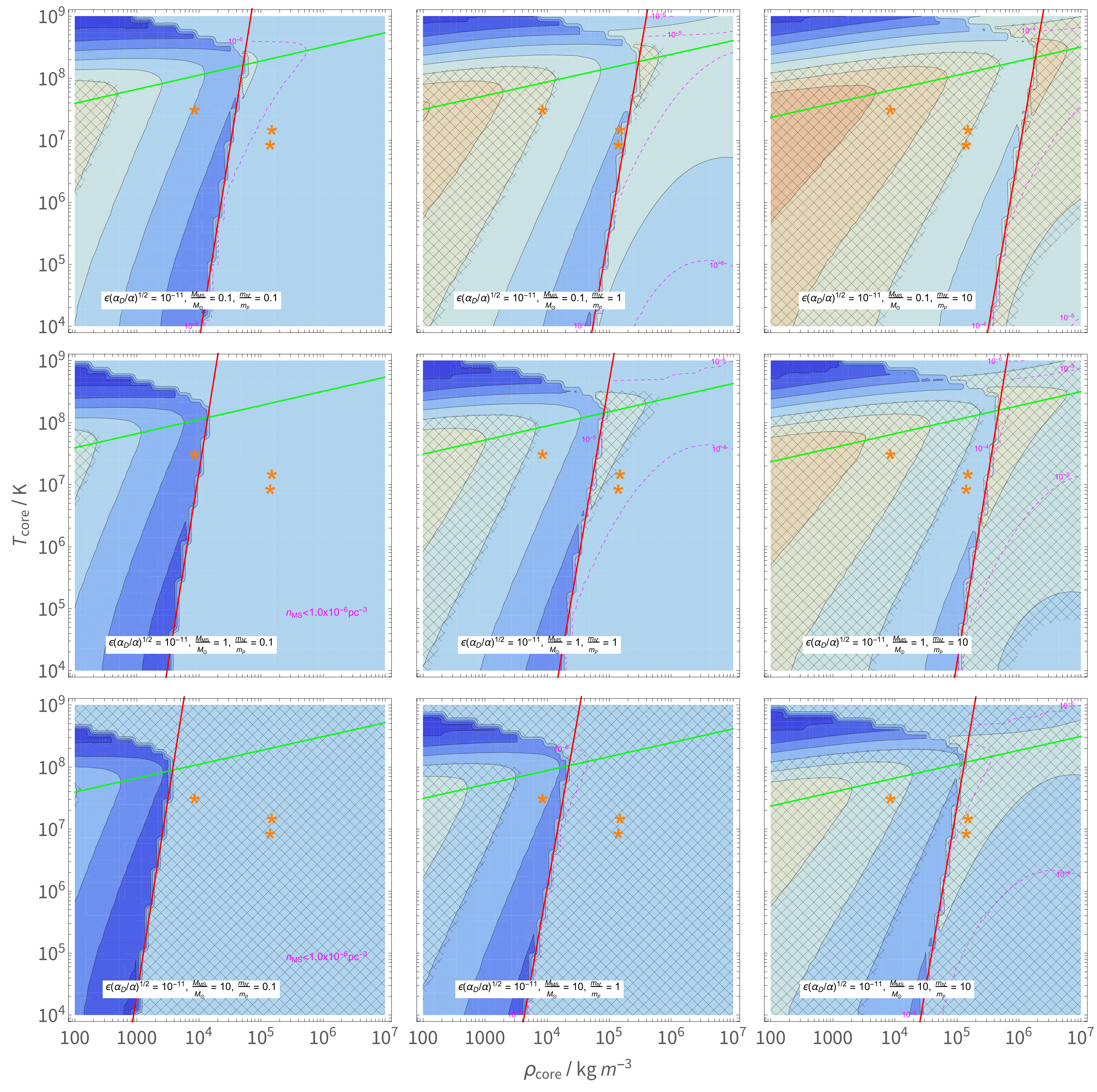}
    \includegraphics[scale=0.662]{figs/numberDensityBar.pdf}
    \caption{Same as \fref{fig:grid10} but for
    $\epsilon \sqrt{\alpha_D/\alpha_{em}} = 10^{-11}$.}
    \label{fig:grid11}
\end{figure*}

\begin{figure*}
    \centering
    \includegraphics[width=0.8 \textwidth]{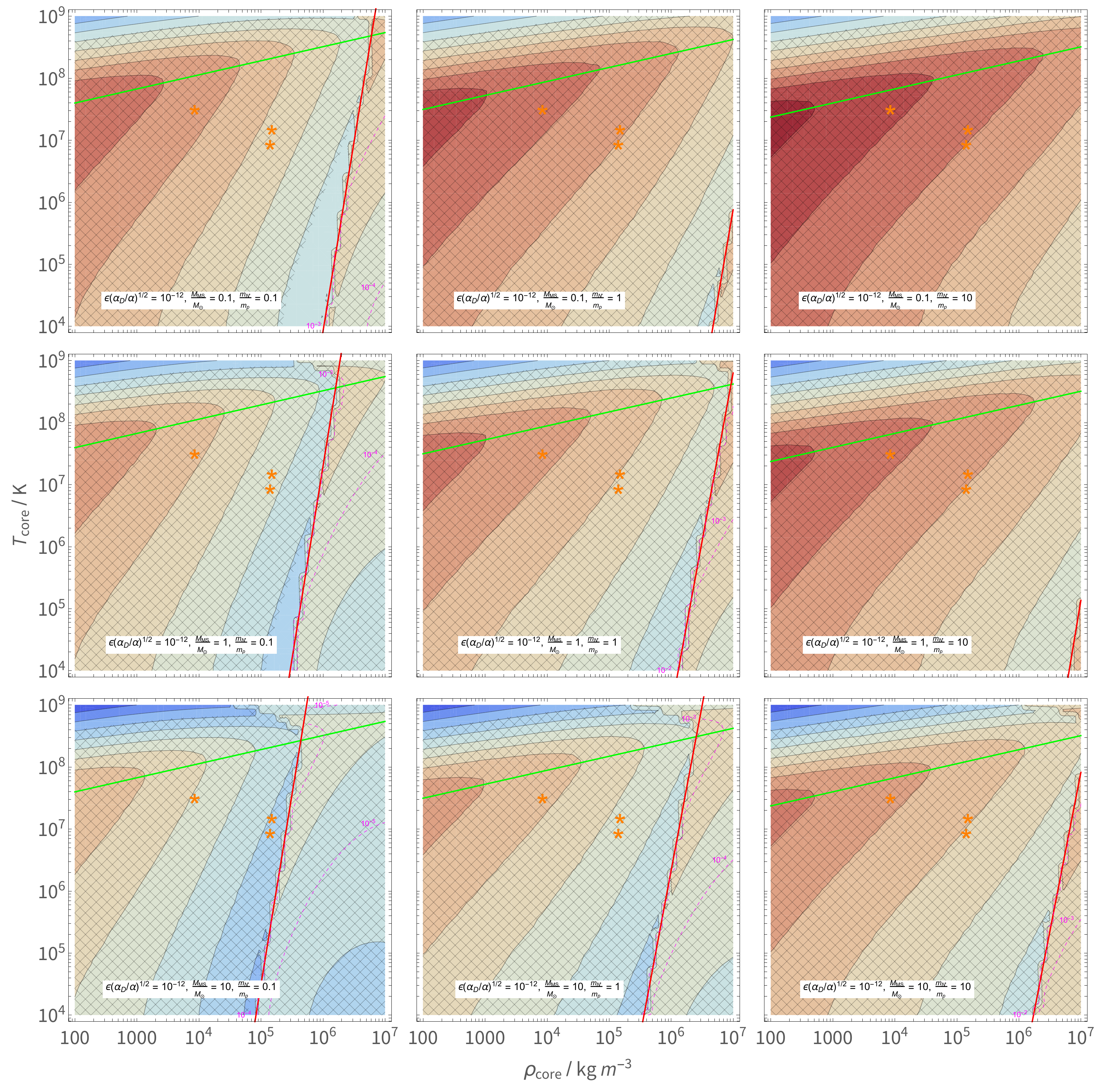}
    \includegraphics[scale=0.662]{figs/numberDensityBar.pdf}
    \caption{Same as \fref{fig:grid10} but for
    $\epsilon \sqrt{\alpha_D/\alpha_{em}} = 10^{-12}$.
    }
    \label{fig:grid12}
\end{figure*}

Let us now understand the importance of the mirror star mass distribution. Inspired by the mass function of regular stars, we will make the simplified assumption that the mirror star mass function follows a single power law:
\begin{equation}
    \label{e.MSMF}
    P(M_{MS}) = A \left( \frac{M_{MS}}{M_\odot} \right)^\beta  \ \ \ \ \mathrm{for}  \ \ \ \  M > M_{MS}^{min}.
\end{equation}
The normalization constant $A$ sets the total mirror star number density, while $\beta$ is expected to be a negative $\mathcal{O}(1)$ number that depends on mirror stellar physics and the dynamics of atomic dark matter in our galaxy and stellar neighborhood. 
$M_{MS}^{min}$ is the \emph{minimum} mass of a mirror star that could be observable. 
This is a useful parameterization if the atomic dark matter sector also contains an equivalent of dark nuclear physics, so that mirror stars actually undergo energy conversion in their cores that drastically prolongs their main sequence lifetime compared to the Kelvin-Helmholz time. In that case, in analogy to the mass of the lightest red dwarf stars in the SM of $\sim 0.1 M_\odot$, there will be some minimum mirror star mass that can initiate ``dark fusion''.\footnote{Depending on details of the dark sector, it is possible that ``mirror brown dwarfs'' contribute significantly to the population of observable objects. We leave this possibility for future investigation.} 

This simple mass function glosses over details at masses much larger than $M_{MS}^{min}$ (for reasons that will be come obvious in a moment) and only has three parameters. The normalization $A$ is what we would like to constrain with observations. In theory, $M_{MS}^{min}$ might be computed from first principles with a knowledge of the hidden sector Lagrangian, but it is for now regarded as a free parameter in our toy mirror star model.  On the other hand, the dependence of $\beta$ on microphysics is so complicated that it has to be regarded, for the present purposes, as unknowable. It is therefore useful to understand the effect of $\beta$ on our $n_{MS}$ constraints.

We repeat the previous analysis with SM-symmetric mirror stars but using the mass function in \eref{e.MSMF} with $\beta$ ranging from $-1$ (flatter than the SM value above red dwarf masses) to $-5$, as well as a delta function mass distribution.  The resulting constraints on $n_{MS}$ for different values of $\epsilon \sqrt{\alpha_D/\alpha_{em}}$ are shown in \fref{f.deltafnbounds} as a function of mirror star (minimum) mass $M_{MS}$. 
In each case, the shading between the dashed curves ($\beta  = - 1$) and the solid curves (delta function mass distribution) represents the effect of the unknown shape of the mirror star mass function. In almost all cases, the effect is relatively minor, at most a factor of 10 or so on the bounds for $n_{MS}$. 
This is perfectly reasonable: the lightest mirror stars are expected to be the most numerous, and when setting constraints, we are in the low-statistics regime where we find the mirror star abundance for which only a handful of mirror stars are expected to be observable. In that case, we would expect most of those nearby mirror stars to be ``red mirror dwarfs'', and the more massive but much rarer 
``blue mirror giants'' are not expected to play a large role in setting our limit. 

This has a very important consequence: for the purpose of setting limits on the abundance of mirror stars from their non-observation, we can take the $\beta \to - \infty$ limit and pretend that all mirror stars are ``red dwarfs''.

\subsection{Constraining mirror star abundance as a function of ``mirror red dwarf'' properties}

We now set constraints on the total local number density of mirror stars as a function of the properties of ``mirror red dwarfs'', i.e. their mass $M_{MS} \equiv M_{MS}^{min}$, radius, core temperature, nuclear constituent mass $m_{N}$, and the mixing parameter $\epsilon \sqrt{\alpha_D/\alpha_{em}}$. As discussed above, in the low-statistics limit the observation of mirror stars is entirely dominated by the most common and lightest kind,  justifying this enormous simplification.

To further reduce the dimensionality of the signal parameter space, we make the additional assumption that mirror red dwarfs, like SM red dwarfs, have a much longer lifetime than the current age of the universe. Therefore, the age of mirror stars today is independent of their actual lifetime. While there is some dependence on the history of mirror matter in our galaxy, we simplify our calculation by setting all mirror stars to have the same representative lifetime (half the age of the universe), which should be a reasonable approximation at our level of precision. 

We also eliminate the mirror star radius as a free parameter by defining $f = \rho_{core}/\rho_{avg}$, the ratio between the mirror star core and average density. For SM stars, $f \sim 1 - 300$, so we fix $f = 100$ in our numerical analysis for simplicity, setting $r_{MS}$ as a function of $M_{MS}$. As we will show below, our final constraints on the mirror star density scale modestly with this unknown ratio, $n_{MS}^{max} \sim \sqrt{f}$, meaning our constraints could be easily adapted to a given understanding of mirror star physics which determines the true value of $f$. 

It is now a straightforward exercise to repeat the analysis of the previous subsection verbatim, for mirror red dwarfs with a given $M_{MS}, T_{core}, \rho_{core}$ and nuclear constituent mass $m_N$, assuming different kinetic mixings. The resulting Gaia bounds on the mirror star number density for $\epsilon \sqrt{\alpha_D/\alpha_{DM}} = 10^{-10}, 10^{-11}$ and $10^{-12}$ are shown in Figs.~\ref{fig:grid10}, \ref{fig:grid11} and \ref{fig:grid12} respectively. 
The coloured contours show maximum allowed $n_{MS}$ as a function of $\rho_{core}$ and $T_{core}$, for mirror red dwarf masses $M_{MS} = 0.1, 1, 10 M_\odot$ (top to bottom) and nuclear constituent masses $m_N = 0.1, 1, 10 m_p$ (left to right). 
In each plot, to the left (right) of the red line, constraints from the optically thin (thick) region provide the strongest constraint. 
Above the green line, converted X-rays from the mirror star core contribute to heating the SM nugget, changing the dependence of the nugget spectrum and magnitude on mirror star parameters. 
The dashed magenta contours in the region where optically thick constraints are dominant indicate the $n_{MS}$ bounds that could be achieved with a background-free search, i.e. our optimistic future sensitivity projection in those parts of parameter space. 
Finally, the three asterisks in each figure indicate $\rho_{core}, T_{core}$ of a typical SM star with mass $0.2, 1, 10 M_\odot$ (in order of increasing core temperature, as a reference.
The constraints vary widely, but at the order-of-magnitude level can  be described reasonably well by an empirical fit function to simple power laws. In the collisional heating-dominated optically thin region, 
\begin{multline}
\label{e.nMSmaxopticallythin}
    n_{MS}^{max} \sim 10\,\textrm{pc}^{-3} \left(\frac{\epsilon}{10^{-12}}\sqrt{\frac{\alpha_D}{\alpha_{em}}}\right)^{-3.33} \left(\frac{\rho_{core}}{10^5\textrm{kg}\,\textrm{m}^{-3}}\right)^{-0.67} \\ \left(\frac{T_{core}}{10^7\textrm{K}}\right)^{0.6} \left(\frac{M_{MS}}{M_\odot}\right)^{-1} \left(\frac{m_N}{m_p}\right)^{0.83}.
\end{multline}
In the high-$T_{core}$ X-ray heating dominated region the bound is roughly
\begin{multline}
\label{e.nMSmaxXray}
    n_{MS}^{max} \sim 350\,\textrm{pc}^{-3}\left(\frac{\epsilon}{10^{-12}}\sqrt{\frac{\alpha_D}{\alpha_{em}}}\right)^{-3.33}\left(\frac{\rho_{core}}{10^3 \textrm{kg}\,\textrm{m}^{-3}}\right) \\
    \left(\frac{T_{core}}{10^8
\textrm{K}}\right)^{-6.67} \left(\frac{M_{MS}}{M_\odot}\right)^{-1}.
\end{multline}
It is possible to understand these two sets of power laws from analytical estimates of the capture and heating rate, using the expressions in~\cite{Curtin:2019ngc}. Assuming that accumulation quickly becomes self-capture dominated, the total number of SM nuclei in the nugget is proportional to $C_{self} \sim v_{esc}^2 R^2$. $v_{esc} \sim \sqrt{M_{MS}/r_{MS}} \sim M_{MS}^{1/3} \rho_{core}^{1/6} f^{-1/6}$ is the mirror star escape velocity, and $R \sim \sqrt{T_{eq}/\rho_{core}}$ is its virial radius, assuming an average equilibrium nugget temperature $T_{eq}$. The heating rate scales as $dP/dV \sim \epsilon^2 \alpha_D n_{SM} \rho_{core} /\sqrt{m_N T_{core}}$ for nuclear scattering, and $dP/dV \sim \epsilon^2 \alpha_D n_{SM} T_{core}^4$ for X-ray heating, where $n_{SM} \sim C_{self}/R^3$ is the number density of captured SM nuclei. The absolute nugget luminosity in the optically thin case then scales as $L_{abs} \sim R^3 dP/dV$, and the corresponding limit on the mirror star number density from Gaia scales as $n_{MS}^{max} \sim L_{abs}^{-3/2}$ based on requiring less than 4 predicted observations above Gaia's apparent luminosity threshold. Putting all this together, we find the following analytical expectation for the approximate scaling of the mirror star constraints:
\begin{equation}
\label{e.nMSmaxestimate}
n_{MS}^{max} \sim 
\frac{\sqrt{f}   \ T_{eq}^{-3/2} }{(\epsilon \sqrt{\alpha_D })^3 M_{MS}}
\left\{
\begin{array}{lll}
\frac{m_N^{3/4} T_{core}^{3/4}}{\rho_{core}^{1/2}} & & \mathrm{(optically\ thin)}\\
\frac{\rho_{core}}{T_{core}^6} & & \mathrm{(X-ray\ heating)}
\end{array}
\right. \ .
\end{equation}
Bremsstrahlung cooling (which has to equal the heating rate) depends on the ionization fraction of the nugget and hence depends exponentially on $T_{eq}$, ensuring the nugget temperature is always close to the ionization energy of SM hydrogen with little dependence on mirror star parameters. 
Therefore, \eref{e.nMSmaxestimate} is in very reasonable agreement with Eqns.~(\ref{e.nMSmaxopticallythin}) and~(\ref{e.nMSmaxXray}), especially given the crudeness of this estimate. \eref{e.nMSmaxestimate} also makes the dependence of our constraints on  $f = \rho_{core}/\rho_{avg}$ explicit. The different dependencies of the limit on $T_{core}$ in the two regions can now also be understood as arising from the radically different temperature dependencies of the respective dominant heating mechanisms.

In each row of Figs.~\ref{fig:grid10}, \ref{fig:grid11} and \ref{fig:grid12} , the total mirror star mass density is linearly related to the mirror star number density by
\begin{equation}
    \rho_{MS} = 127 \cdot \frac{n_{MS}}{\textrm{pc}^{-3}} \cdot \rho_{DM} \cdot \left(\frac{M_{MS}}{M_\odot} \right),
\end{equation}
where we assume a local measured DM mass density of $\rho_{DM} = 0.3~\mathrm{GeV}/\mathrm{cm}^{-3}$~\cite{Bovy:2012tw,Benito:2019ngh}.
To facilitate interpretation of these $n_{MS}$ constraints in terms of their local contribution to the dark matter mass density, the hatched regions indicate where in parameter space the Gaia constraints are \emph{weaker} than the constraint that $\rho_{MS} < \rho_{DM}$. In some figures there are no such regions, so we also show the $\rho_{MS} = 0.01 \rho_{DM}$ contour as a thick black curve in each figure to allow for an intuitive conversion between $n_{MS}$ and $\rho_{MS}$ constraints.

For $\epsilon \sqrt{\alpha_D/\alpha_{em}} \gtrsim 10^{-11}$, there are large regions of mirror star parameter space where the direct observational constraints from  Gaia are stronger than the strongest existing constraint that $\rho_{MS} \leq \rho_{DM}$. For example, for mirror sectors that are similar to the SM, such as in the mirror twin higgs model, the middle column of \fref{fig:grid10} shows that optically thin mirror stars can be constrained to be at most a percent-level or even smaller fraction of the local DM, depending on the exact properties of the mirror stars.  On the other hand, for the parameters we have considered, Gaia is not sensitive to relevant mirror star densities when $\epsilon \sqrt{\alpha_D/\alpha_{em}} \lesssim 10^{-12}$.

\section{Discussion and Conclusion}
\label{s.conclusion}

In this paper we demonstrated how to conduct the first direct search for dark matter using Gaia data. Mirror stars, predicted in theories of atomic dark matter,  could conceivably show up in Gaia as very dim stars with significantly different colour than main sequence stars or white dwarfs. 
We calculated the simplified emission spectra for mirror stars with  widely varying  properties, which demonstrates that mirror stars are likely to appear in two distinct regions of the HR diagram, corresponding to optically thin or optically thick SM nuggets of captured interstellar matter.
We then conducted a provisional search for optically thin mirror stars using their motivated signal region, and identify 14 candidates closer than 100 pc in the Gaia dataset. 
For the three highest-quality candidates we were able to assemble  pseudo-spectra by combining the Gaia data with measurements from other stellar catalogues. 
Our simple attempts to clearly identify these candidates as either white dwarfs or mirror stars was inconclusive, but they are likely just white dwarfs that 
are particularly dim and red because of a combination of high mass and old age, and possibly also because of dust extinction. Even so, our analysis demonstrates how to conduct a search for mirror stars using  Gaia data despite the wide range of possible dark sector parameters and unknown aspects of mirror stellar physics. This will serve as blueprint for a realistic search with a more complete calculation of the SM nugget emissions in the future.

Non-observation of mirror stars in Gaia would set strong constraints on atomic dark matter models. We compute projected Gaia exclusions on their local number density  beyond the requirement that their contribution to the local DM mass density does not exceed the measured value of $\rho_{DM} \approx 0.3 \mathrm{GeV}/\mathrm{cm}^3$. 
Since this depends mostly on the bolometric magnitude of the mirror star SM photon signal, these projections should give an accurate idea of Gaia's sensitivity even with the simplified calculation of mirror star emissions.
It is noteworthy that constraints can be derived without detailed knowledge of the mirror star mass distribution, since the constraint will be dominated by the least massive and most numerous "mirror red dwarfs" that can sustain ``dark nuclear fusion'' (assuming the dark sector includes some energy conversion mechanism to ignite the cores of mirror stars). 
We can therefore set constraints on the mirror red dwarf mass, core temperature, core density and nuclear constituent mass. Across many regions of mirror star parameter space, their contribution to the local DM mass fraction would be  constrained to the percent-level or below.

Our analysis sets the stage for a number of important future investigations, the most urgent of which is obtaining a more realistic calculation for the optical/thermal emission spectra of the captured SM nuggets. 
This includes solving for the structure of optically thick nuggets by taking convection and opacity into account, the inclusion of additional emission processes beyond free-free or black body, and taking into account the presence of metals heavier than helium in the interstellar medium, which may alter the shape of nugget emission spectra in a highly distinctive way.
This would also be a necessary ingredient for a more careful optically thick mirror star search, which would either take the known distribution of white dwarfs in the HR diagram into account as a ``background", or use spectral analysis to distinguish them from optically thick mirror stars.

We identified that mirror star constraints are dominated by ``mirror red dwarfs''. Unlike the detailed mass-dependent abundance of the whole mirror star population, the properties 
of the lightest-possible mirror star for a given dark sector may actually be discernable from the microphysics, i.e. the dark sector Lagrangian.
If certain details of the ``dark nuclear physics'' can be constrained or parameterized this would be a fascinating avenue for future investigation.

\begin{figure}
    \centering
    \includegraphics[width=0.4\textwidth]{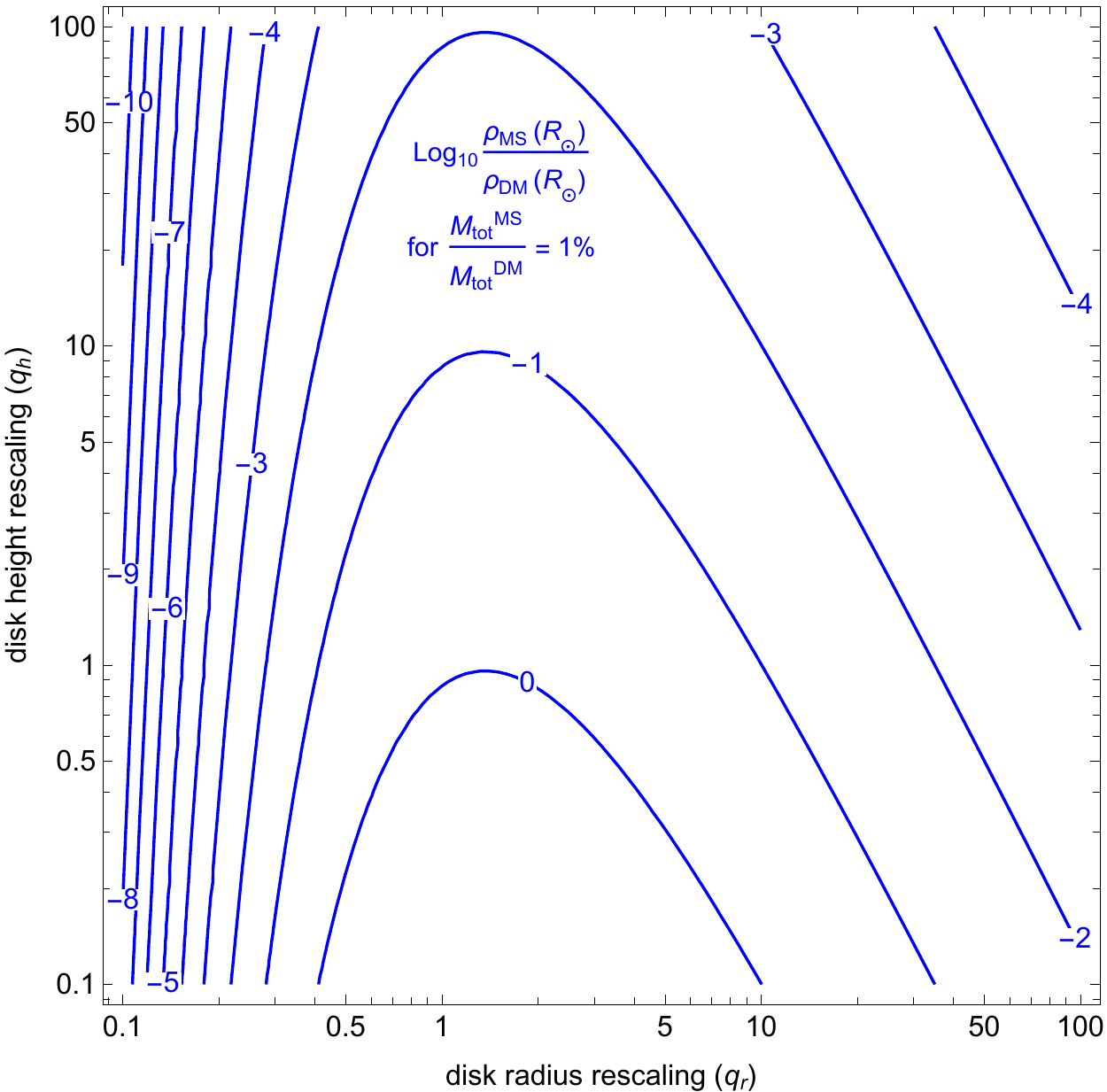}
    \caption{Expected local mirror star mass density relative to local DM density of $\rho_{DM}(R_\odot) = 0.3 \mathrm{GeV}/\mathrm{cm}^3$ if 1\% of DM in our milky way is in the form of mirror stars, and assuming they are distributed in a double-exponential galactic disk distribution with radius and height rescaled by factors $q_r$ and $q_h$ relative to the visible milky way disk~(see~\cite{Winch:2020cju}).
    }
    \label{f.MSdensity}
\end{figure}

Our study also focused exclusively on the visible signal of main sequence mirror stars. However, it is possible that mirror stellar relics provide an even stronger signal~\cite{Curtin:2020tkm}, due to their higher initial core temperatures and densities. Since it is already known how to relate the properties of stellar relics to fundamental microphysical parameters~\cite{Reisenegger:2015crq,Hippert:2021fch}, this may also allow us to constrain the dark sector Lagrangian more directly. 

The richness of possible dark sector dynamics makes the observational and theoretical study of dark complexity extremely challenging.
This is simply demonstrated by trying to place our Gaia constraints on the local mirror star density in the context of other observational searches for exotic astrophysical phenomena. 
In Ref.~\cite{Winch:2020cju}, the sensitivity of the Rubin telescope/LSST microlensing survey to mirror stars in a dark-disk-like configurations was studied. 
The possible shape of the dark disk was parameterized by rescaling the radius and height of the visible milky way disk, and projected sensitivities to the fraction of dark matter in our milky way that is made up of mirror stars in that dark disk were derived. Sensitivities to sub-percent-level galaxy-wide DM fractions are possible depending on the shape of the dark disk.
Crucially, the local DM fraction, and hence the observation of mirror stars in Gaia, depends very sensitively on the properties of the dark disk as well. 
As demonstrated in \fref{f.MSdensity}, a percent-level DM fraction of mirror stars in our galaxy can lead to local mirror star DM mass fractions that range from 100\% to smaller than 1\%, depending on the radius and height of the dark disk. 
Non-observation of mirror stars in one survey does not mean they might not be detected in another one.

On the flip side, the richness of possible dark sector dynamics means that any discovery would be the harbinger of a true golden age of dark sector particle and astro-physics. For example, if  Gaia actually observed mirror stars, it could measure their proper motions and provide detailed information about the local mirror star distribution.
Together with microlensing surveys~\cite{Winch:2020cju}, gravitational wave searches for mirror neutron stars~\cite{Hippert:2021fch}, direct detection experiments~\cite{Kaplan:2011yj,Fan:2013bea,Chacko:2005pe} and other searches this would paint an incredibly vivid picture of the dark world that we might share our universe with. Such an incredible opportunity  must motivate us to understand dark complexity in all its possible forms. 

\vspace{10mm}
\emph{Acknowledgements:} 

We thank Simon Blouin for useful discussion on the properties of white dwarfs. 
The research of AH, JS and DC is supported in part by a Discovery Grant from the Natural Sciences and Engineering Research Council of Canada, and by the Canada Research Chair program. The research of CDM is  supported by a separate NSERC Discovery Grant.  JS also acknowledges support from the University of Toronto Faculty of Arts and Science postdoctoral fellowship. 
This publication makes use of data products from: (1) the ESA mission {\it GAIA} processed by the {\it Gaia}
Data Processing and Analysis Consortium; (2) the Two Micron All Sky Survey (a UMass/JPL project funded by NASA and the NSF); 
(3) WISE (a joint UCLA/JPL project  funded by NASA) and NeoWISE (a JPL project funded by NASA's Planetary Science Division; 
(4) the DENIS project, funded by the European Commission and grants in several member states; 
(5) the Pan-STARRS1 project and its archives, made possible by contributions from numerous U.S. and European universities and networks and funding from NASA, NSF, as well as the Gordon and Betty Moore Foundation. 

\bibliographystyle{JHEP}
\bibliography{References}

\providecommand{\href}[2]{#2}\begingroup\raggedright\begin{thebibliography}{10}

\bibitem{2016A&A...595A...1G}
{Gaia Collaboration}, T.~{Prusti}, J.~H.~J. {de Bruijne}, A.~G.~A. {Brown},
  A.~{Vallenari}, C.~{Babusiaux} et~al., \emph{{The Gaia mission}},
  \href{https://doi.org/10.1051/0004-6361/201629272}{\emph{\aap} {\bfseries
  595} (Nov., 2016) A1}, [\href{https://arxiv.org/abs/1609.04153}{{\ttfamily
  1609.04153}}].

\bibitem{Sanderson:2014apa}
R.~E. Sanderson, A.~Helmi and D.~W. Hogg, \emph{{Action-space clustering of
  tidal streams to infer the Galactic potential}},
  \href{https://doi.org/10.1017/S1743921313006388}{\emph{IAU Symp.} {\bfseries
  298} (2014) 207}, [\href{https://arxiv.org/abs/1404.6534}{{\ttfamily
  1404.6534}}].

\bibitem{Ostdiek:2019gnb}
B.~Ostdiek, L.~Necib, T.~Cohen, M.~Freytsis, M.~Lisanti, S.~Garrison-Kimmel
  et~al., \emph{{Cataloging Accreted Stars within $Gaia$ DR2 using Deep
  Learning}}, \href{https://doi.org/10.1051/0004-6361/201936866}{\emph{Astron.
  Astrophys.} {\bfseries 636} (2020) A75},
  [\href{https://arxiv.org/abs/1907.06652}{{\ttfamily 1907.06652}}].

\bibitem{Schutz:2017tfp}
K.~Schutz, T.~Lin, B.~R. Safdi and C.-L. Wu, \emph{{Constraining a Thin Dark
  Matter Disk with Gaia}},
  \href{https://doi.org/10.1103/PhysRevLett.121.081101}{\emph{Phys. Rev. Lett.}
  {\bfseries 121} (2018) 081101},
  [\href{https://arxiv.org/abs/1711.03103}{{\ttfamily 1711.03103}}].

\bibitem{Necib:2018igl}
L.~Necib, M.~Lisanti, S.~Garrison-Kimmel, A.~Wetzel, R.~Sanderson, P.~F.
  Hopkins et~al., \emph{{Under the Firelight: Stellar Tracers of the Local Dark
  Matter Velocity Distribution in the Milky Way}},
  \href{https://arxiv.org/abs/1810.12301}{{\ttfamily 1810.12301}}.

\bibitem{Necib:2018iwb}
L.~Necib, M.~Lisanti and V.~Belokurov, \emph{{Inferred Evidence For Dark Matter
  Kinematic Substructure with SDSS-Gaia}},
  \href{https://arxiv.org/abs/1807.02519}{{\ttfamily 1807.02519}}.

\bibitem{1960BAN....15...45O}
J.~H. {Oort}, \emph{{Note on the determination of K$_{z}$ and on the mass
  density near the Sun.}}, {\emph{{BAIN}} {\bfseries 15} (Feb., 1960) 45}.

\bibitem{2015ApJ...814...13M}
C.~F. {McKee}, A.~{Parravano} and D.~J. {Hollenbach}, \emph{{Stars, Gas, and
  Dark Matter in the Solar Neighborhood}},
  \href{https://doi.org/10.1088/0004-637X/814/1/13}{\emph{\apj} {\bfseries 814}
  (Nov., 2015) 13}, [\href{https://arxiv.org/abs/1509.05334}{{\ttfamily
  1509.05334}}].

\bibitem{Curtin:2019lhm}
D.~Curtin and J.~Setford, \emph{{How To Discover Mirror Stars}},
  \href{https://doi.org/10.1016/j.physletb.2020.135391}{\emph{Phys. Lett. B}
  {\bfseries 804} (2020) 135391},
  [\href{https://arxiv.org/abs/1909.04071}{{\ttfamily 1909.04071}}].

\bibitem{Curtin:2019ngc}
D.~Curtin and J.~Setford, \emph{{Signatures of Mirror Stars}},
  \href{https://doi.org/10.1007/JHEP03(2020)041}{\emph{JHEP} {\bfseries 03}
  (2020) 041}, [\href{https://arxiv.org/abs/1909.04072}{{\ttfamily
  1909.04072}}].

\bibitem{Goldberg:1986nk}
H.~Goldberg and L.~J. Hall, \emph{{A New Candidate for Dark Matter}},
  \href{https://doi.org/10.1016/0370-2693(86)90731-8}{\emph{Phys. Lett. B}
  {\bfseries 174} (1986) 151}.

\bibitem{Kaplan:2009de}
D.~E. Kaplan, G.~Z. Krnjaic, K.~R. Rehermann and C.~M. Wells, \emph{{Atomic
  Dark Matter}},
  \href{https://doi.org/10.1088/1475-7516/2010/05/021}{\emph{JCAP} {\bfseries
  05} (2010) 021}, [\href{https://arxiv.org/abs/0909.0753}{{\ttfamily
  0909.0753}}].

\bibitem{Kaplan:2011yj}
D.~E. Kaplan, G.~Z. Krnjaic, K.~R. Rehermann and C.~M. Wells, \emph{{Dark
  Atoms: Asymmetry and Direct Detection}},
  \href{https://doi.org/10.1088/1475-7516/2011/10/011}{\emph{JCAP} {\bfseries
  10} (2011) 011}, [\href{https://arxiv.org/abs/1105.2073}{{\ttfamily
  1105.2073}}].

\bibitem{Cline:2012is}
J.~M. Cline, Z.~Liu and W.~Xue, \emph{{Millicharged Atomic Dark Matter}},
  \href{https://doi.org/10.1103/PhysRevD.85.101302}{\emph{Phys. Rev. D}
  {\bfseries 85} (2012) 101302},
  [\href{https://arxiv.org/abs/1201.4858}{{\ttfamily 1201.4858}}].

\bibitem{Cline:2013pca}
J.~M. Cline, Z.~Liu, G.~Moore and W.~Xue, \emph{{Scattering properties of dark
  atoms and molecules}},
  \href{https://doi.org/10.1103/PhysRevD.89.043514}{\emph{Phys. Rev. D}
  {\bfseries 89} (2014) 043514},
  [\href{https://arxiv.org/abs/1311.6468}{{\ttfamily 1311.6468}}].

\bibitem{Fan:2013yva}
J.~Fan, A.~Katz, L.~Randall and M.~Reece, \emph{{Double-Disk Dark Matter}},
  \href{https://doi.org/10.1016/j.dark.2013.07.001}{\emph{Phys. Dark Univ.}
  {\bfseries 2} (2013) 139--156},
  [\href{https://arxiv.org/abs/1303.1521}{{\ttfamily 1303.1521}}].

\bibitem{Fan:2013tia}
J.~Fan, A.~Katz, L.~Randall and M.~Reece, \emph{{Dark-Disk Universe}},
  \href{https://doi.org/10.1103/PhysRevLett.110.211302}{\emph{Phys. Rev. Lett.}
  {\bfseries 110} (2013) 211302},
  [\href{https://arxiv.org/abs/1303.3271}{{\ttfamily 1303.3271}}].

\bibitem{Fan:2013bea}
J.~Fan, A.~Katz and J.~Shelton, \emph{{Direct and indirect detection of
  dissipative dark matter}},
  \href{https://doi.org/10.1088/1475-7516/2014/06/059}{\emph{JCAP} {\bfseries
  06} (2014) 059}, [\href{https://arxiv.org/abs/1312.1336}{{\ttfamily
  1312.1336}}].

\bibitem{Cyr-Racine:2013fsa}
F.-Y. Cyr-Racine, R.~de~Putter, A.~Raccanelli and K.~Sigurdson,
  \emph{{Constraints on Large-Scale Dark Acoustic Oscillations from
  Cosmology}}, \href{https://doi.org/10.1103/PhysRevD.89.063517}{\emph{Phys.
  Rev. D} {\bfseries 89} (2014) 063517},
  [\href{https://arxiv.org/abs/1310.3278}{{\ttfamily 1310.3278}}].

\bibitem{Rosenberg:2017qia}
E.~Rosenberg and J.~Fan, \emph{{Cooling in a Dissipative Dark Sector}},
  \href{https://doi.org/10.1103/PhysRevD.96.123001}{\emph{Phys. Rev. D}
  {\bfseries 96} (2017) 123001},
  [\href{https://arxiv.org/abs/1705.10341}{{\ttfamily 1705.10341}}].

\bibitem{Ghalsasi:2017jna}
A.~Ghalsasi and M.~McQuinn, \emph{{Exploring the astrophysics of dark atoms}},
  \href{https://doi.org/10.1103/PhysRevD.97.123018}{\emph{Phys. Rev. D}
  {\bfseries 97} (2018) 123018},
  [\href{https://arxiv.org/abs/1712.04779}{{\ttfamily 1712.04779}}].

\bibitem{Gresham:2018anj}
M.~I. Gresham, H.~K. Lou and K.~M. Zurek, \emph{{Astrophysical Signatures of
  Asymmetric Dark Matter Bound States}},
  \href{https://doi.org/10.1103/PhysRevD.98.096001}{\emph{Phys. Rev. D}
  {\bfseries 98} (2018) 096001},
  [\href{https://arxiv.org/abs/1805.04512}{{\ttfamily 1805.04512}}].

\bibitem{Essig:2018pzq}
R.~Essig, S.~D. Mcdermott, H.-B. Yu and Y.-M. Zhong, \emph{{Constraining
  Dissipative Dark Matter Self-Interactions}},
  \href{https://doi.org/10.1103/PhysRevLett.123.121102}{\emph{Phys. Rev. Lett.}
  {\bfseries 123} (2019) 121102},
  [\href{https://arxiv.org/abs/1809.01144}{{\ttfamily 1809.01144}}].

\bibitem{Alvarez:2019nwt}
G.~Alvarez and H.-B. Yu, \emph{{Astrophysical probes of inelastic dark matter
  with a light mediator}},
  \href{https://doi.org/10.1103/PhysRevD.101.043002}{\emph{Phys. Rev. D}
  {\bfseries 101} (2020) 043002},
  [\href{https://arxiv.org/abs/1911.11114}{{\ttfamily 1911.11114}}].

\bibitem{Cline:2021itd}
J.~M. Cline, \emph{{Dark atoms and composite dark matter}},  in \emph{{Les
  Houches summer school on Dark Matter}}, 8, 2021,
  \href{https://arxiv.org/abs/2108.10314}{{\ttfamily 2108.10314}}.

\bibitem{Cyr-Racine:2021alc}
F.-Y. Cyr-Racine, F.~Ge and L.~Knox, \emph{{A Symmetry of Cosmological
  Observables, and a High Hubble Constant as an Indicator of a Mirror World
  Dark Sector}},  \href{https://arxiv.org/abs/2107.13000}{{\ttfamily
  2107.13000}}.

\bibitem{Blinov:2021mdk}
N.~Blinov, G.~Krnjaic and S.~W. Li, \emph{{Towards a Realistic Model of Dark
  Atoms to Resolve the Hubble Tension}},
  \href{https://arxiv.org/abs/2108.11386}{{\ttfamily 2108.11386}}.

\bibitem{Chacko:2005pe}
Z.~Chacko, H.-S. Goh and R.~Harnik, \emph{{The Twin Higgs: Natural electroweak
  breaking from mirror symmetry}},
  \href{https://doi.org/10.1103/PhysRevLett.96.231802}{\emph{Phys. Rev. Lett.}
  {\bfseries 96} (2006) 231802},
  [\href{https://arxiv.org/abs/hep-ph/0506256}{{\ttfamily hep-ph/0506256}}].

\bibitem{Barbieri:2005ri}
R.~Barbieri, T.~Gregoire and L.~J. Hall, \emph{{Mirror world at the large
  hadron collider}},  \href{https://arxiv.org/abs/hep-ph/0509242}{{\ttfamily
  hep-ph/0509242}}.

\bibitem{Chacko:2005vw}
Z.~Chacko, Y.~Nomura, M.~Papucci and G.~Perez, \emph{{Natural little hierarchy
  from a partially goldstone twin Higgs}},
  \href{https://doi.org/10.1088/1126-6708/2006/01/126}{\emph{JHEP} {\bfseries
  01} (2006) 126}, [\href{https://arxiv.org/abs/hep-ph/0510273}{{\ttfamily
  hep-ph/0510273}}].

\bibitem{Chacko:2016hvu}
Z.~Chacko, N.~Craig, P.~J. Fox and R.~Harnik, \emph{{Cosmology in Mirror Twin
  Higgs and Neutrino Masses}},
  \href{https://doi.org/10.1007/JHEP07(2017)023}{\emph{JHEP} {\bfseries 07}
  (2017) 023}, [\href{https://arxiv.org/abs/1611.07975}{{\ttfamily
  1611.07975}}].

\bibitem{Craig:2016lyx}
N.~Craig, S.~Koren and T.~Trott, \emph{{Cosmological Signals of a Mirror Twin
  Higgs}}, \href{https://doi.org/10.1007/JHEP05(2017)038}{\emph{JHEP}
  {\bfseries 05} (2017) 038},
  [\href{https://arxiv.org/abs/1611.07977}{{\ttfamily 1611.07977}}].

\bibitem{Chacko:2018vss}
Z.~Chacko, D.~Curtin, M.~Geller and Y.~Tsai, \emph{{Cosmological Signatures of
  a Mirror Twin Higgs}},
  \href{https://doi.org/10.1007/JHEP09(2018)163}{\emph{JHEP} {\bfseries 09}
  (2018) 163}, [\href{https://arxiv.org/abs/1803.03263}{{\ttfamily
  1803.03263}}].

\bibitem{Chacko:2021vin}
Z.~Chacko, D.~Curtin, M.~Geller and Y.~Tsai, \emph{{Direct Detection of Mirror
  Matter in Twin Higgs Models}},
  \href{https://arxiv.org/abs/2104.02074}{{\ttfamily 2104.02074}}.

\bibitem{GarciaGarcia:2015pnn}
I.~Garcia~Garcia, R.~Lasenby and J.~March-Russell, \emph{{Twin Higgs Asymmetric
  Dark Matter}},
  \href{https://doi.org/10.1103/PhysRevLett.115.121801}{\emph{Phys. Rev. Lett.}
  {\bfseries 115} (2015) 121801},
  [\href{https://arxiv.org/abs/1505.07410}{{\ttfamily 1505.07410}}].

\bibitem{Foot:2002iy}
R.~Foot and S.~Mitra, \emph{{Ordinary atom mirror atom bound states: A New
  window on the mirror world}},
  \href{https://doi.org/10.1103/PhysRevD.66.061301}{\emph{Phys. Rev. D}
  {\bfseries 66} (2002) 061301},
  [\href{https://arxiv.org/abs/hep-ph/0204256}{{\ttfamily hep-ph/0204256}}].

\bibitem{Foot:2003jt}
R.~Foot and R.~R. Volkas, \emph{{Was ordinary matter synthesized from mirror
  matter? An Attempt to explain why Omega(Baryon) approximately equal to 0.2
  Omega(Dark)}}, \href{https://doi.org/10.1103/PhysRevD.68.021304}{\emph{Phys.
  Rev. D} {\bfseries 68} (2003) 021304},
  [\href{https://arxiv.org/abs/hep-ph/0304261}{{\ttfamily hep-ph/0304261}}].

\bibitem{Berezhiani:2003xm}
Z.~Berezhiani, \emph{{Mirror world and its cosmological consequences}},
  \href{https://doi.org/10.1142/S0217751X04020075}{\emph{Int. J. Mod. Phys. A}
  {\bfseries 19} (2004) 3775--3806},
  [\href{https://arxiv.org/abs/hep-ph/0312335}{{\ttfamily hep-ph/0312335}}].

\bibitem{Foot:1999hm}
R.~Foot, \emph{{Have mirror stars been observed?}},
  \href{https://doi.org/10.1016/S0370-2693(99)00230-0}{\emph{Phys. Lett. B}
  {\bfseries 452} (1999) 83--86},
  [\href{https://arxiv.org/abs/astro-ph/9902065}{{\ttfamily
  astro-ph/9902065}}].

\bibitem{Foot:2000vy}
R.~Foot, A.~Y. Ignatiev and R.~R. Volkas, \emph{{Physics of mirror photons}},
  \href{https://doi.org/10.1016/S0370-2693(01)00228-3}{\emph{Phys. Lett. B}
  {\bfseries 503} (2001) 355--361},
  [\href{https://arxiv.org/abs/astro-ph/0011156}{{\ttfamily
  astro-ph/0011156}}].

\bibitem{Foot:2003eq}
R.~Foot, \emph{{Experimental implications of mirror matter - type dark
  matter}}, \href{https://doi.org/10.1142/S0217751X04020087}{\emph{Int. J. Mod.
  Phys. A} {\bfseries 19} (2004) 3807--3818},
  [\href{https://arxiv.org/abs/astro-ph/0309330}{{\ttfamily
  astro-ph/0309330}}].

\bibitem{Foot:2004pa}
R.~Foot, \emph{{Mirror matter-type dark matter}},
  \href{https://doi.org/10.1142/S0218271804006449}{\emph{Int. J. Mod. Phys. D}
  {\bfseries 13} (2004) 2161--2192},
  [\href{https://arxiv.org/abs/astro-ph/0407623}{{\ttfamily
  astro-ph/0407623}}].

\bibitem{Bansal:2021dfh}
S.~Bansal, J.~H. Kim, C.~Kolda, M.~Low and Y.~Tsai, \emph{{Mirror Twin Higgs
  Cosmology: Constraints and a Possible Resolution to the $H_0$ and $S_8$
  Tensions}},  \href{https://arxiv.org/abs/2110.04317}{{\ttfamily 2110.04317}}.

\bibitem{Berezhiani:2005vv}
Z.~Berezhiani, S.~Cassisi, P.~Ciarcelluti and A.~Pietrinferni,
  \emph{{Evolutionary and structural properties of mirror star MACHOs}},
  \href{https://doi.org/10.1016/j.astropartphys.2005.10.002}{\emph{Astropart.
  Phys.} {\bfseries 24} (2006) 495--510},
  [\href{https://arxiv.org/abs/astro-ph/0507153}{{\ttfamily
  astro-ph/0507153}}].

\bibitem{Holdom:1985ag}
B.~Holdom, \emph{{Two U(1)'s and Epsilon Charge Shifts}},
  \href{https://doi.org/10.1016/0370-2693(86)91377-8}{\emph{Phys. Lett. B}
  {\bfseries 166} (1986) 196--198}.

\bibitem{Vogel:2013raa}
H.~Vogel and J.~Redondo, \emph{{Dark Radiation constraints on minicharged
  particles in models with a hidden photon}},
  \href{https://doi.org/10.1088/1475-7516/2014/02/029}{\emph{JCAP} {\bfseries
  02} (2014) 029}, [\href{https://arxiv.org/abs/1311.2600}{{\ttfamily
  1311.2600}}].

\bibitem{Gherghetta:2019coi}
T.~Gherghetta, J.~Kersten, K.~Olive and M.~Pospelov, \emph{{Evaluating the
  price of tiny kinetic mixing}},
  \href{https://doi.org/10.1103/PhysRevD.100.095001}{\emph{Phys. Rev. D}
  {\bfseries 100} (2019) 095001},
  [\href{https://arxiv.org/abs/1909.00696}{{\ttfamily 1909.00696}}].

\bibitem{Winch:2020cju}
H.~Winch, J.~Setford, J.~Bovy and D.~Curtin, \emph{{Using LSST Microlensing to
  Constrain Dark Compact Objects in Spherical and Disk Configurations}},
  \href{https://arxiv.org/abs/2012.07136}{{\ttfamily 2012.07136}}.

\bibitem{Hippert:2021fch}
M.~Hippert, J.~Setford, H.~Tan, D.~Curtin, J.~Noronha-Hostler and N.~Yunes,
  \emph{{Mirror Neutron Stars}},
  \href{https://arxiv.org/abs/2103.01965}{{\ttfamily 2103.01965}}.

\bibitem{Adams08}
F.~C. {Adams}, \emph{{Stars in other universes: stellar structure with
  different fundamental constants}},
  \href{https://doi.org/10.1088/1475-7516/2008/08/010}{\emph{{JCAP}} {\bfseries
  2008} (Aug., 2008) 010}, [\href{https://arxiv.org/abs/0807.3697}{{\ttfamily
  0807.3697}}].

\bibitem{Curtin:2020tkm}
D.~Curtin and J.~Setford, \emph{{Direct Detection of Atomic Dark Matter in
  White Dwarfs}}, \href{https://doi.org/10.1007/JHEP03(2021)166}{\emph{JHEP}
  {\bfseries 03} (2021) 166},
  [\href{https://arxiv.org/abs/2010.00601}{{\ttfamily 2010.00601}}].

\bibitem{Gaidau:2021vyr}
C.~Gaidau and J.~Shelton, \emph{{Singularities in the gravitational capture of
  dark matter through long-range interactions}},
  \href{https://doi.org/10.1088/1475-7516/2022/01/016}{\emph{JCAP} {\bfseries
  01} (2022) 016}, [\href{https://arxiv.org/abs/2110.02234}{{\ttfamily
  2110.02234}}].

\bibitem{DeRocco:2022rze}
W.~DeRocco, M.~Galanis and R.~Lasenby, \emph{{Dark matter scattering in
  astrophysical media: collective effects}},
  \href{https://doi.org/10.1088/1475-7516/2022/05/015}{\emph{JCAP} {\bfseries
  05} (2022) 015}, [\href{https://arxiv.org/abs/2201.05167}{{\ttfamily
  2201.05167}}].

\bibitem{Freedman14_Opacities}
R.~S. {Freedman}, J.~{Lustig-Yaeger}, J.~J. {Fortney}, R.~E. {Lupu}, M.~S.
  {Marley} and K.~{Lodders}, \emph{{Gaseous Mean Opacities for Giant Planet and
  Ultracool Dwarf Atmospheres over a Range of Metallicities and Temperatures}},
  \href{https://doi.org/10.1088/0067-0049/214/2/25}{\emph{The Astrophysical
  Journal Supplement} {\bfseries 214} (Oct., 2014) 25},
  [\href{https://arxiv.org/abs/1409.0026}{{\ttfamily 1409.0026}}].

\bibitem{2018A&A...616A...4E}
D.~W. {Evans}, M.~{Riello}, F.~{De Angeli}, J.~M. {Carrasco}, P.~{Montegriffo},
  C.~{Fabricius} et~al., \emph{{Gaia Data Release 2. Photometric content and
  validation}}, \href{https://doi.org/10.1051/0004-6361/201832756}{\emph{\aap}
  {\bfseries 616} (Aug., 2018) A4},
  [\href{https://arxiv.org/abs/1804.09368}{{\ttfamily 1804.09368}}].

\bibitem{Bovy:2012tw}
J.~Bovy and S.~Tremaine, \emph{{On the local dark matter density}},
  \href{https://doi.org/10.1088/0004-637X/756/1/89}{\emph{Astrophys. J.}
  {\bfseries 756} (2012) 89},
  [\href{https://arxiv.org/abs/1205.4033}{{\ttfamily 1205.4033}}].

\bibitem{Benito:2019ngh}
M.~Benito, A.~Cuoco and F.~Iocco, \emph{{Handling the Uncertainties in the
  Galactic Dark Matter Distribution for Particle Dark Matter Searches}},
  \href{https://doi.org/10.1088/1475-7516/2019/03/033}{\emph{JCAP} {\bfseries
  03} (2019) 033}, [\href{https://arxiv.org/abs/1901.02460}{{\ttfamily
  1901.02460}}].

\bibitem{2018A&A...616A...1G}
{Gaia Collaboration}, A.~G.~A. {Brown}, A.~{Vallenari}, T.~{Prusti}, J.~H.~J.
  {de Bruijne}, C.~{Babusiaux} et~al., \emph{{Gaia Data Release 2. Summary of
  the contents and survey properties}},
  \href{https://doi.org/10.1051/0004-6361/201833051}{\emph{\aap} {\bfseries
  616} (Aug., 2018) A1}, [\href{https://arxiv.org/abs/1804.09365}{{\ttfamily
  1804.09365}}].

\bibitem{Lindegren:2018cgr}
L.~Lindegren et~al., \emph{{Gaia Data Release 2: The astrometric solution}},
  \href{https://doi.org/10.1051/0004-6361/201832727}{\emph{Astron. Astrophys.}
  {\bfseries 616} (2018) A2},
  [\href{https://arxiv.org/abs/1804.09366}{{\ttfamily 1804.09366}}].

\bibitem{2020arXiv200907277G}
P.~{Gandhi}, D.~A.~H. {Buckley}, P.~{Charles}, S.~{Hodgkin}, S.~{Scaringi},
  C.~{Knigge} et~al., \emph{{Astrometric excess noise in Gaia DR2 and the
  search for X-ray emitting binaries}}, {\emph{arXiv e-prints} (Sept., 2020)
  arXiv:2009.07277}, [\href{https://arxiv.org/abs/2009.07277}{{\ttfamily
  2009.07277}}].

\bibitem{gentile2019gaia}
N.~P. Gentile~Fusillo, P.-E. Tremblay, B.~T. G{\"a}nsicke, C.~J. Manser,
  T.~Cunningham, E.~Cukanovaite et~al., \emph{A gaia data release 2 catalogue
  of white dwarfs and a comparison with sdss}, {\emph{Monthly Notices of the
  Royal Astronomical Society} {\bfseries 482} (2019) 4570--4591}.

\bibitem{Kong21_WDs_in_LAMOST}
X.~{Kong} and A.-L. {Luo}, \emph{{Identification of White Dwarf from Gaia EDR3
  via Spectra from LAMOST DR7}}, {\emph{arXiv e-prints} (Oct., 2021)
  arXiv:2110.00002}, [\href{https://arxiv.org/abs/2110.00002}{{\ttfamily
  2110.00002}}].

\bibitem{chambers2019panstarrs1}
K.~C. Chambers, E.~A. Magnier, N.~Metcalfe, H.~A. Flewelling, M.~E. Huber,
  C.~Z. Waters et~al., \emph{The pan-starrs1 surveys},  2019.

\bibitem{1993ASPC...52...21D}
E.~R. {Deul}, \emph{{DENIS---DEep Near Infrared Survey of the Southern Sky}},
  in \emph{Astronomical Data Analysis Software and Systems II} (R.~J.
  {Hanisch}, R.~J.~V. {Brissenden} and J.~{Barnes}, eds.), vol.~52 of
  \emph{Astronomical Society of the Pacific Conference Series}, p.~21, Jan.,
  1993.

\bibitem{Kirkpatric14:allwise}
J.~D. {Kirkpatrick}, A.~{Schneider}, S.~{Fajardo-Acosta}, C.~R. {Gelino}, G.~N.
  {Mace}, E.~L. {Wright} et~al., \emph{{The AllWISE Motion Survey and the Quest
  for Cold Subdwarfs}},
  \href{https://doi.org/10.1088/0004-637X/783/2/122}{\emph{\apj} {\bfseries
  783} (Mar., 2014) 122}, [\href{https://arxiv.org/abs/1402.0661}{{\ttfamily
  1402.0661}}].

\bibitem{2006AJ....131.1163S}
M.~F. {Skrutskie}, R.~M. {Cutri}, R.~{Stiening}, M.~D. {Weinberg},
  S.~{Schneider}, J.~M. {Carpenter} et~al., \emph{{The Two Micron All Sky
  Survey (2MASS)}}, \href{https://doi.org/10.1086/498708}{\emph{{AJ}}
  {\bfseries 131} (Feb., 2006) 1163--1183}.

\bibitem{GATOR}
``{GATOR} catalog search facility.''
  \url{https://irsa.ipac.caltech.edu/applications/Gator}.

\bibitem{2010A&A...523A..48J}
C.~{Jordi}, M.~{Gebran}, J.~M. {Carrasco}, J.~{de Bruijne}, H.~{Voss},
  C.~{Fabricius} et~al., \emph{{Gaia broad band photometry}},
  \href{https://doi.org/10.1051/0004-6361/201015441}{\emph{\aap} {\bfseries
  523} (Nov., 2010) A48}, [\href{https://arxiv.org/abs/1008.0815}{{\ttfamily
  1008.0815}}].

\bibitem{2018MNRAS.479L.102C}
L.~{Casagrande} and D.~A. {VandenBerg}, \emph{{On the use of Gaia magnitudes
  and new tables of bolometric corrections}},
  \href{https://doi.org/10.1093/mnrasl/sly104}{\emph{Monthly Notices of the
  Royal Astronomical Society} {\bfseries 479} (Sept., 2018) L102--L107},
  [\href{https://arxiv.org/abs/1806.01953}{{\ttfamily 1806.01953}}].

\bibitem{2018ascl.soft11001S}
{STScI Development Team}, \emph{{synphot: Synthetic photometry using Astropy}},
   Nov., 2018.

\bibitem{Torres21_GaiaWDs}
S.~{Torres}, A.~{Rebassa-Mansergas}, M.~E. {Camisassa} and R.~{Raddi},
  \emph{{The Gaia DR2 halo white dwarf population: the luminosity function,
  mass distribution, and its star formation history}},
  \href{https://doi.org/10.1093/mnras/stab079}{\emph{MNRAS} {\bfseries 502}
  (Apr., 2021) 1753--1767}, [\href{https://arxiv.org/abs/2101.03341}{{\ttfamily
  2101.03341}}].

\bibitem{holberg2006calibration}
J.~Holberg and P.~Bergeron, \emph{Calibration of synthetic photometry using da
  white dwarfs}, {\emph{The Astronomical Journal} {\bfseries 132} (2006) 1221}.

\bibitem{blouin2018new}
S.~Blouin, P.~Dufour and N.~F. Allard, \emph{A new generation of cool white
  dwarf atmosphere models. i. theoretical framework and applications to dz
  stars}, {\emph{The Astrophysical Journal} {\bfseries 863} (2018) 184}.

\bibitem{bedard2020spectral}
A.~B{\'e}dard, P.~Bergeron, P.~Brassard and G.~Fontaine, \emph{On the spectral
  evolution of hot white dwarf stars. i. a detailed model atmosphere analysis
  of hot white dwarfs from sdss dr12}, {\emph{The Astrophysical Journal}
  {\bfseries 901} (2020) 93}.

\bibitem{tremblay2011improved}
P.-E. Tremblay, P.~Bergeron and A.~Gianninas, \emph{An improved spectroscopic
  analysis of da white dwarfs from the sloan digital sky survey data release
  4}, {\emph{The Astrophysical Journal} {\bfseries 730} (2011) 128}.

\bibitem{BergeronSite}
``Synthetic colors and evolutionary sequences of hydrogen- and
  helium-atmosphere white dwarfs.''
  \url{https://www.astro.umontreal.ca/\~bergeron/CoolingModels/}.

\bibitem{Kroupa:2000iv}
P.~Kroupa, \emph{{On the variation of the initial mass function}},
  \href{https://doi.org/10.1046/j.1365-8711.2001.04022.x}{\emph{Mon. Not. Roy.
  Astron. Soc.} {\bfseries 322} (2001) 231},
  [\href{https://arxiv.org/abs/astro-ph/0009005}{{\ttfamily
  astro-ph/0009005}}].

\bibitem{Paxton_2010}
B.~Paxton, L.~Bildsten, A.~Dotter, F.~Herwig, P.~Lesaffre and F.~Timmes,
  \emph{Modules for experiments in stellar astrophysics (mesa)},
  \href{https://doi.org/10.1088/0067-0049/192/1/3}{\emph{The Astrophysical
  Journal Supplement Series} {\bfseries 192} (Dec, 2010) 3}.

\bibitem{Paxton_2013}
B.~Paxton, M.~Cantiello, P.~Arras, L.~Bildsten, E.~F. Brown, A.~Dotter et~al.,
  \emph{Modules for experiments in stellar astrophysics (mesa): Planets,
  oscillations, rotation, and massive stars},
  \href{https://doi.org/10.1088/0067-0049/208/1/4}{\emph{The Astrophysical
  Journal Supplement Series} {\bfseries 208} (Aug, 2013) 4}.

\bibitem{Paxton_2015}
B.~Paxton, P.~Marchant, J.~Schwab, E.~B. Bauer, L.~Bildsten, M.~Cantiello
  et~al., \emph{Modules for experiments in stellar astrophysics (mesa):
  Binaries, pulsations, and explosions},
  \href{https://doi.org/10.1088/0067-0049/220/1/15}{\emph{The Astrophysical
  Journal Supplement Series} {\bfseries 220} (Sep, 2015) 15}.

\bibitem{Paxton:2017eie}
B.~Paxton et~al., \emph{{Modules for Experiments in Stellar Astrophysics
  (MESA): Convective Boundaries, Element Diffusion, and Massive Star
  Explosions}},
  \href{https://doi.org/10.3847/1538-4365/aaa5a8}{\emph{Astrophys. J. Suppl.}
  {\bfseries 234} (2018) 34},
  [\href{https://arxiv.org/abs/1710.08424}{{\ttfamily 1710.08424}}].

\bibitem{Paxton_2019}
B.~Paxton, R.~Smolec, J.~Schwab, A.~Gautschy, L.~Bildsten, M.~Cantiello et~al.,
  \emph{Modules for experiments in stellar astrophysics (mesa): Pulsating
  variable stars, rotation, convective boundaries, and energy conservation},
  \href{https://doi.org/10.3847/1538-4365/ab2241}{\emph{The Astrophysical
  Journal Supplement Series} {\bfseries 243} (Jul, 2019) 10}.

\bibitem{Reisenegger:2015crq}
A.~Reisenegger and F.~S. Zepeda, \emph{{Order-of-magnitude physics of neutron
  stars}}, \href{https://doi.org/10.1140/epja/i2016-16052-y}{\emph{Eur. Phys.
  J. A} {\bfseries 52} (2016) 52},
  [\href{https://arxiv.org/abs/1511.08813}{{\ttfamily 1511.08813}}].

\end{thebibliography}\endgroup

\end{document}